\documentclass{article}
\usepackage{graphicx}
\usepackage{float}
\usepackage[toc,page]{appendix}
\usepackage{amsmath, amsthm, latexsym}
\usepackage{amssymb}
\usepackage{subfigure}
\usepackage{pdfsync}
\usepackage{hyperref}

\usepackage{diagbox}
 \usepackage[utf8]{inputenc}
\begin{document}

\vspace*{3cm} 
\vspace{5mm}

 \textbf{\Large Polar Jet Stream Fluctuations in an Energy\\
  Balance Model}\\

\textbf{\normalsize  \textbf{\normalsize Cord Perillo\footnote{Department of Mathematics and Interdisciplinary Research Institute for the Sciences, California State University, Northridge, Northridge, CA 91330-8313. Email: cord.perillo@csun.edu.}, David Klein\footnote{Department of Mathematics and Interdisciplinary Research Institute for the Sciences, California State University, Northridge, Northridge, CA 91330-8313. Email: david.klein@csun.edu. Orcid ID:  https://orcid.org/0000-0003-1964-4378}, Rabia Djellouli  \footnote{Department of Mathematics and Interdisciplinary Research Institute for the Sciences, California State University, Northridge, Northridge, CA 91330-8313. Email: rabia.djellouli@csun.edu.}}}
\\

\vspace{4mm} \parbox{11cm}{{\small \textbf{Abstract.} We investigate the effect of increased longwave radiative forcing (a proxy for increased greenhouse gas concentration) on the zonally averaged location of the eddy-driven jet stream in a latitude dependent, two-layer Energy Balance Model. The model includes separate terms for atmospheric and surface albedos, and takes into account reflections of shortwave radiation between the surface and atmospheric layers. We introduce the notion of a cloud factor function, which depends on temperature gradients, to simulate the eddy-driven jet. An increase in longwave radiative forcing initially results in a poleward movement of the jet stream's mean latitude, but as the forcing increases, the location of the jet stream becomes quasi-periodic and its mean location moves equatorward.
}\\


 }\\
\vspace{6cm}
\pagebreak

\setlength{\voffset}{-0.75in}
\setlength{\textheight}{650pt}

\section{Introduction}

The atmosphere and the ocean stabilize Earth's climate from uneven solar insolation by transporting heat from the equator to the poles. Energy balance models (EBMs), first introduced by Budyko \cite{budyko}, Sellers \cite{sellers}, include heat transport terms that reproduce zonally and annually averaged temperature profiles from this transport.  These idealized climate models have been extensively studied (e.g. North \cite{North75, North81, North83}) and a wide range of modifications and additional forcings have been introduced in order to provide insights into causal relationships of components of Earth's climate, for example, \cite{ Held, Domm, Ikeda, jentsch, Merlis, Lindzen, Stocker2, soder, Bonetti, KE} among many other studies.\\

In this paper, we use an energy balance model to investigate the dynamics of the polar jet stream of an aqua-planet in response to increasing greenhouse gas concentrations, with a focus on the role of cloud fraction and albedo.  Both observations and climate model studies indicate that the general circulation pattern of the atmosphere is altered by anthropogenic warming, e.g., \cite{Barnes,Francis,  hu, karamperidou, Lu, Manney, yin, zhou}.  Among these are two studies that employed EBMs to investigate the link between shifts of the midlatitude storm tracks to the shifts of the Hadley cell edge:  Mbengue and Schneider \cite{Mbengue} (hereafter MS18) and and Siler, Roe, Armour  \cite{siler} (hereafter SRA18).\\

MS18 \cite{Mbengue} defined the storm track in a one layer EBM as the latitude of maximum absolute value of the temperature gradient.  In that model, the diffusion coefficient was increased within the Hadley cell, relative to the diffusion coefficient outside the cell, and the Hadley cell edge (or terminus) was interactive and also depended on the convective lapse rate $\gamma$ in the tropics, which was treated as a parameter.  The model predicts that storm tracks shift in tandem as the Hadley cell edge is moved poleward  by decreasing $\gamma$.  Their results also indicate that strengthening meridional temperature gradient at the Hadley cell terminus can reduce the distance between the Hadley cell edge and the storm tracks, resulting in storm tracks that do not parallel shifts of the Hadley cell terminus.\\ 

SRA18 \cite{siler} studied a single layer Moist Energy Balance perturbation model.  Assuming a reference climate determined by reanalysis or averages of climate models, their perturbation model determines a change in temperature and in evaporation minus precipitation, $E-P$, as a function of latitude, from forcings such as increased greenhouse gas concentrations.  The extratropical latitude of the minimum value of $E-P$ serves as the proxy for mid-latitude storm tracks.  In the case of spatially uniform radiative forcing, SRA18 \cite{siler} found that down-gradient energy transport implies a poleward expansion of the subtropics where $E - P > 0$, and a poleward shift in the extratropical minimum of $E - P$, consistent with a poleward shift of storm-track latitudes.\\  

The idealized model considered in this paper is a latitude dependent, two-layer energy balance model that includes separate terms for atmospheric and surface albedos, and takes into account reflections of shortwave radiation between the surface and atmospheric layers, and includes heat diffusion terms for each layer.  The novel feature of our model is what we refer to as a ``cloud factor function'', a function which depends on temperature gradients, and which dynamically simulates the eddy-driven or polar jet stream.
 More specifically, at any fixed time, the cloud factor function, $\mathcal{C}_f(\theta)$, is a dimensionless quantity that represents the fraction of the zonally averaged planetary albedo at latitude $\theta$ attributable to clouds. 
We use it to construct the atmospheric albedo as a function of latitude at each time step in our model (see Section \ref{cloudfactorsection} below).\\ 
 
 The thermal wind equations link the horizontal temperature gradient to the polar-front jet and suggest that the location of the jet  may be identified with the location of the maximum magnitude of the extratropical temperature gradient; this proxy was utilized in MS18 \cite{Mbengue}. Similar to MS18 \cite{Mbengue}, we interpret the latitude where this occurs as the averaged location of the eddy-driven jet,
 and define our cloud factor function to achieve a maximum value at that location at each time step in our numerical scheme. This allows us to track location of the jet
 as it moves dynamically until the system reaches equilibrium.\\

 We must point out that the Hadley cell edge is not interactive in our model. We hold it fixed at 30$^\circ$ latitude in our numerical experiments. However, this location can easily be modiflied, and the qualitative behavior of our model is robust with respect to this location. Despite this constraint, our model identifies a driver of jet stream fluctuations which has the potential to be incorporated into more complex climate models that include Hadley cell dynamics.\\

The cloud factor function, $\mathcal{C}_f(\theta)$, is constructed so that its minimum corresponds to the Hadley cell boundary and so that the lowest extratropical latitude, at which $\mathcal{C}_f(\theta)$ reaches a prescribed maximum value, identifies the mean location of the polar jet stream. We assume that the cloud factor function is given by a cubic Hermite spline. This spline is defined by specified values of the cloud factor function at four latitudes --- the equator, the Hadley cell edge, the polar jet stream, and the pole. These values are fixed, but one of the latitudes --- the polar jet stream latitude --- is a function of the temperature gradient. There is thus one degree of freedom in the cloud factor function. The location of the jet is determined by the gradient of the average of the atmospheric and surface temperatures.\\

This paper is organized as follows.  Section \ref{Model Description} is divided into subsections that describe the components of our model, including standard forcings, but which focus primarily on the couplings between the cloud factor function and the surface and atmospheric albedos. We also describe how the latitude, where the maximum magnitude of the temperature gradient occurs at each time step of our computations, alters the cloud factor function for the next time step.  Section \ref{results} describes the results of numerical experiments for changes in the location of the eddy-driven jet
as radiative forcing increases, such as from increasing greenhouse gas concentrations.  In Section \ref{conclude}, we compare the behavior of our model with other investigations of jet stream response to increasing greenhouse gas concentrations and offer concluding remarks.  In addition, there are three appendices.  Appendix \ref{sec:A} gives an explicit formula for the cloud factor function; Appendix \ref{program} provides a concise description of the numerical scheme used in our computations; and Appendix \ref{quasiperiodic} displays output data.

\section{Model Description}\label{Model Description}

Our EBM consists of an ocean covered surface layer and an overlying atmospheric layer.  Throughout, we let $x=\sin\phi$, where $\phi$ is latitude\footnote{This formula assumes that $\phi$ is measured in radians. Later, in the context of temperature gradients, it will be calculated as  $x=\sin(\pi\theta/180)$ where $\theta$ is given in degrees.}, so that $-1\leq x\leq 1$, but because our aqua-planet is symmetrical, we will generally display data only for the northern hemisphere, $0\leq x\leq 1$.\\  

 Let $T_s$ and $T_a$ represent the zonally averaged temperatures of the surface and atmosphere respectively, expressed in degrees Celsius. Here, $T_a$ is a measure of the free tropospheric temperature, say at 500 hPa, but as in \cite{rose} we express it as an equivalent surface air temperature, assuming a constant lapse rate (depending on the value of parameters used in the model)\footnote{In particular, we will vary the parameter $A_\text{out}$ in Eq. \eqref{eq:odeatm} to simulate changes in greenhouse gas concentrations.}. The time evolution of the  temperatures are solutions to coupled differential equations of the form,

\begin{subequations}
\begin{eqnarray}
C_{a}\dfrac{\partial T_{a}}{\partial t} &=& F_{atm}^{\downarrow}+F_{up} - F_{out} - \dfrac{1}{2\pi a^{2}}\dfrac{d \mathcal{H}_{a}}{dx}\label{eq:ebm2dima}
\\
C_{s}\dfrac{\partial T_{s}}{\partial t} &=& F_{ground}^{\downarrow}  - F_{up} - \dfrac{1}{2\pi a^{2}}\dfrac{d \mathcal{H}_{s}}{dx},
\label{eq:ebm2dimb}
\end{eqnarray}
\end{subequations}\\
where $a$ is the radius of Earth, $C_{a}, C_{s}$ are respectively specific heats of the atmosphere and surface, $F_{out}$ is the longwave radiative heat flux to space, and $F_{up}$ is the net flux of longwave radiation, latent heat, and sensible heat from the ocean to the atmosphere.  The last terms in each equation represent meridional diffusive heat transport (given explicitly in Eqs \eqref{eq:odeatm} and \eqref{eq:odesurface} below). Although multiple processes are involved in heat transport and although they vary across regions and time scales, Stone (1978) \cite{Stone} demonstrated that the magnitude of the annual mean total meridional heat transport is insensitive to the details of dynamics of the atmosphere-ocean system. \\

As described below, these terms will be chosen to match the corresponding terms in the two layer energy balance model of Rose and Marshall \cite{rose, rose1}. By contrast, the remaining two terms, $F_{atm}^{\downarrow}$ and $F_{ground}^{\downarrow}$, in Eqs. \eqref{eq:ebm2dima} and \eqref{eq:ebm2dimb} represent incoming solar radiation flux and both depend on the atmospheric albedo, $\alpha_{a}$, and ground albedo, $\alpha_{g}$.\\

 To model the dependence of $F_{atm}^{\downarrow}$ and $F_{ground}^{\downarrow}$ on $\alpha_{a}$, and  $\alpha_{g}$, we follow Qu and Hall \cite{qu} and Donohoe and Battisti \cite{donohoe}.  We assume an atmospheric layer within which the radiation undergoes three processes: reflection by a factor $\alpha_{a}$, transmission by a factor $\mathcal{T}_{sw}$ (the transmissivity of shortwave radiation), and absorption by a factor $A_{sw} = 1 - \alpha_{a} - \mathcal{T}_{sw}$.\\

Summing up the infinite number of transmissions and reflections between the atmosphere, the ground, the shortwave flux abosrbed by the ground $F_{ground}^{\downarrow}$ and the shortwave flux absorbed by the atmosphere $F_{atm}^{\downarrow}$ and the radiative flux to space from the top of the atmosphere are given by,

\begin{equation}
F_{ground}^{\downarrow} = \dfrac{(1 - \alpha_{g})\mathcal{T}_{sw}}{(1 - \alpha_{a}\alpha_{g})}\dfrac{S_{0}s(x)}{4}
\end{equation}

\begin{equation}
F_{atm}^{\downarrow} = (1 - \alpha_{a} - \mathcal{T}_{sw})(1 + \dfrac{\alpha_{g} \mathcal{T}_{sw}}{1 - \alpha_{a} \alpha_{g}})\dfrac{S_{0}s(x)}{4},
\end{equation}

\begin{equation}\label{TOA}
F_{TOA}^{\uparrow} = (\alpha_{a} + \dfrac{\mathcal{T}_{sw}^{2} \alpha_{g}}{(1 - \alpha_{a} \alpha_{g})})\dfrac{S_{0}s(x)}{4}
\end{equation}
where $s(x)$ is the annual weight function for incoming solar radiation (dimensionless, unit global mean) which, following \cite{rose, rose1}, is given in terms of the second order Legendre polynomial $P_{2}(x)$ as,
\begin{equation}
s(x) = 1 + s_{2}P_{2}(x), 
\label{eq:sx}
\end{equation}
with $s_{2}=-0.48$.\\ 

We note that $F_{ground}^{\downarrow} + F_{atm}^{\downarrow}+ F_{TOA}^{\uparrow} = S_{0}s(x)/4$, that is, the sum of the various components of absorbed and reflected radiation equals the total quantity of incoming solar radiation. Eq. \eqref{TOA} does not appear in our EBM, but it shows that the planetary albedo can be identified as,

\begin{equation}
\alpha_{p} = \alpha_{p,atm} + \alpha_{p,ground} = \alpha_{a} + \mathcal{T}_{e}\alpha_{g}
\label{eq:sumplanalbedo}
\end{equation}

 where
\begin{equation}
\mathcal{T}_{e} = \dfrac{\mathcal{T}_{sw}^{2}}{1 - \alpha_{a} \alpha_{g}},
\end{equation}

 and $\mathcal{T}_{e}\alpha_{g}$ can be considered as the contribution from the ground albedo $\alpha_{g}$ to the planetary albedo modulated by the interactions with the atmosphere.

\subsection{Cloud Factor Function}\label{cloudfactorsection}
 In order to assign latitudinal values to the ground and atmospheric albedos, $\alpha_{g}$ and $\alpha_{a}$, we first introduce a \emph{cloud factor function}, $\mathcal{C}_f = \mathcal{C}_f(\theta, \hat{\theta}(t))$.  At time $t$,  $\mathcal{C}_f$ is a function of latitude $\theta$ and of the location $\hat{\theta}(t)$ of the maximum of the absolute value of the temperature gradient. The function $\mathcal{C}_f$  is related to the zonally averaged albedo at latitude $\theta$ attributable to clouds (see Eq.\eqref{eq:atmalb} below).   
 The cloud factor function, whose general features are motivated by Figure \ref{fig:zonalmeancloud}, is explained in detail in Subsection \ref{subsectjet} below.\\
 

It is difficult to measure cloud cover in the polar regions due to a number of factors, including thin and low lying clouds and polar conditions that create an unusual amount of near surface hazes and fogs \cite{curry}. Cloud fraction in global climate models and atmospheric reanalyses vary widely \cite{ipcc}, and clouds are among the main sources of uncertainty in modeling the Arctic climate \cite{LiuY}. Because of these problems, there is an uncertainty in cloud cover over the polar regions. Vavrus et al. \cite{vavrus} conclude maximum cloudiness occurs over open water in the summer time with cloud fraction values of $81\%$. Palm et al. \cite{palm} agree that maximum cloudiness occurs over open water in the summer time but report model cloud fraction values of $90\%$ or more.  Both conclude that the average polar cloud fraction is increasing as the sea ice extent has been decreasing.\\

In \cite{Noris} and references therein, Norris examined climate variability and found a positive cloud feedback on sea surface temperatures (SST), in the North Pacific during the boreal summer, where increased cloud amount acts to cool the ocean by decreasing surface insolation, and decreased SST favors greater marine stratiform cloudiness amount.  This suggests a steep drop in temperature associated with high cloud cover. In addition, Figures 1, 2, and 3 in SRA18 \cite{siler} indicate maximum precipitation at the minimum of $E-P$ with high values of precipitation poleward.  This suggests high cloud cover poleward of the jet location.

\begin{figure}[H]
\includegraphics[width=1\textwidth]{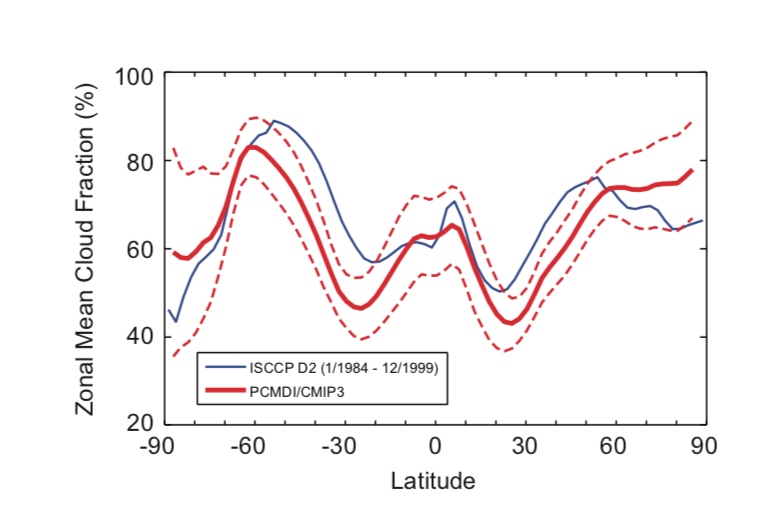}
\caption[Zonal Mean Cloud Fraction]{Zonal mean cloud fraction from CMIP3 models and compared to observations (International Satellite Cloud Climatology Project, ISCCP).\footnotemark}
\label{fig:zonalmeancloud}
\end{figure}
\footnotetext{Figure from Stocker 2022 edition \cite{stocker}.}

\begin{figure}[H]
\center
\includegraphics[width=0.7\textwidth]{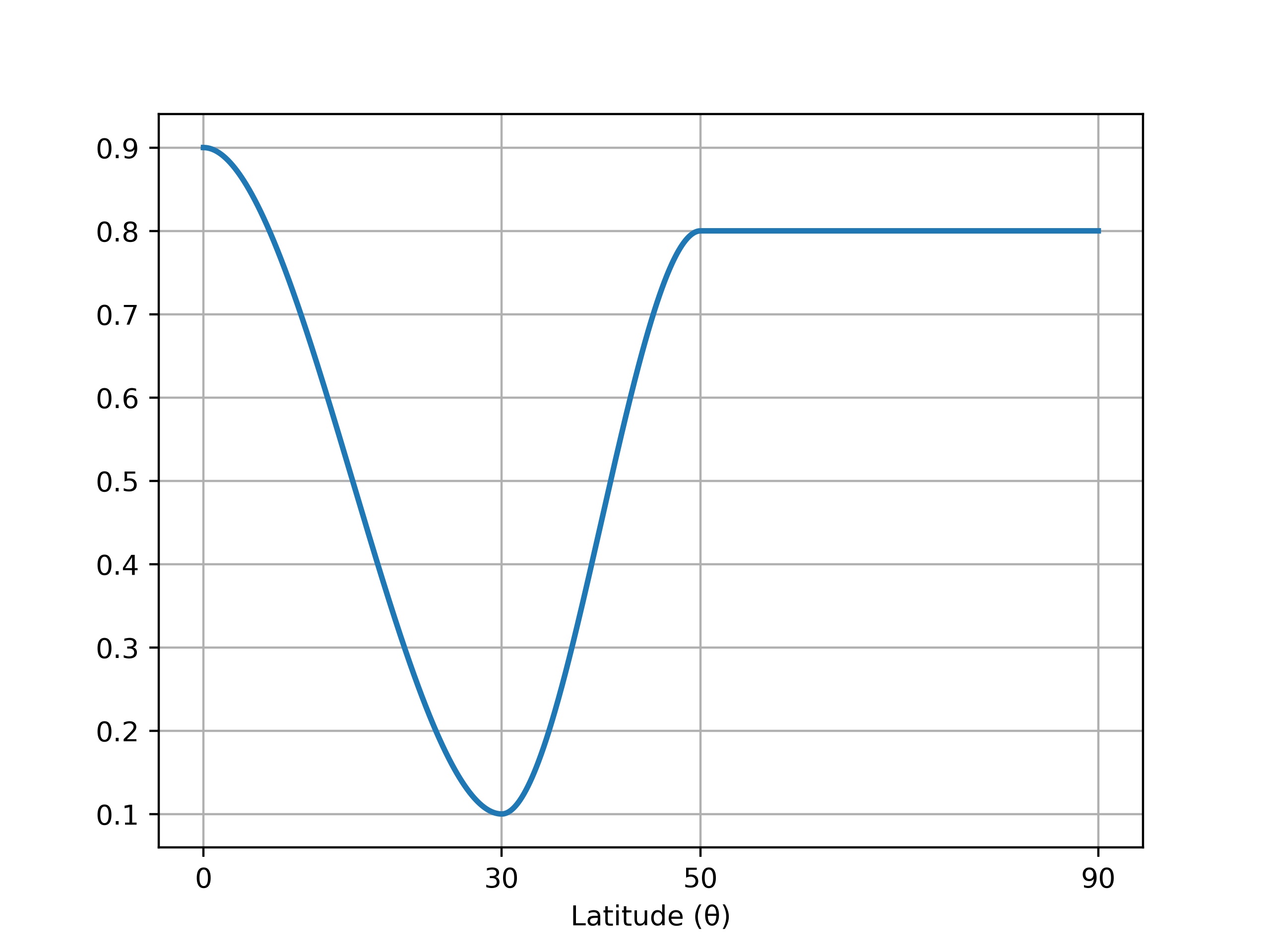}
\caption[Cubic Spline Cloud Factor Three]{Cubic Hermite spline cloud factor $\mathcal{C}_f$ plotted as a function of latitude from equator to pole with the first extratropical maximum $\hat{\theta} = 50^\circ$ latitude.  In general, the location of the the first extratropical maximum is interactive and varies in time. See equation \ref{eq:cfc} in Appendix \ref{sec:A}.}
\label{fig:cfs3}
\end{figure}

Taking these findings into consideration, we construct $\mathcal{C}_f(\theta, \hat{\theta})$ through the use of cubic Hermite splines.\footnote{Cubic Hermite splines are continuously differentiable at all points, including juncture points.  We note that the use of linear splines instead of cubic splines results in qualititatively similar final results.}  The cloud factor function is incorporated into our climate model as described in Section \ref{subsectjet}.  The graph of the cloud factor function is initially constrained to take extremal values  at 0$^\circ$,  30$^\circ$, 50$^\circ$, and 90$^\circ$ latitude, the locations of the equator, the Hadley cell edge, the polar jet stream, and the pole.
Specifically, the coordinates are $(0, 0.9), (30, 0.1), (50, 0.8)$ and $(90,0.8)$ so as to represent high cloudiness at the equator as well as poleward of 50$^\circ$ degrees, and low cloudiness at 30$^\circ$ degrees. However, as we explain in Sect. \ref{subsectjet}, the graph will change with the time steps in the numerical runs of our model.   A sample graph is shown in Figure \ref{fig:cfs3}.  We note that our numerical experiments exhibit the same qualitative behavior, as we describe in this paper, even when the cloud factor function is modified so that the cloud cover varies in the region poleward of the jet
or takes a different constant value in that region. \\

As pointed out in \cite{Mbengue1}, the Southern Hemisphere polar jet is located at $50^\circ$ latitude. So this is a plausible choice for an initial location of the polar jet prior to radiative forcings that we will impose.  We note that the EBM of SAR18 \cite{siler} locates the initial northern hemisphere minimum value of $E-P$ (evaporation minus precipitation and the proxy in that EBM for storm track location) above $60^\circ$ latitude (see Figures 2f and 3b in \cite{siler}).  

\subsection{Albedo Functions}

\begin{figure}[H]
\includegraphics[width=1\textwidth]{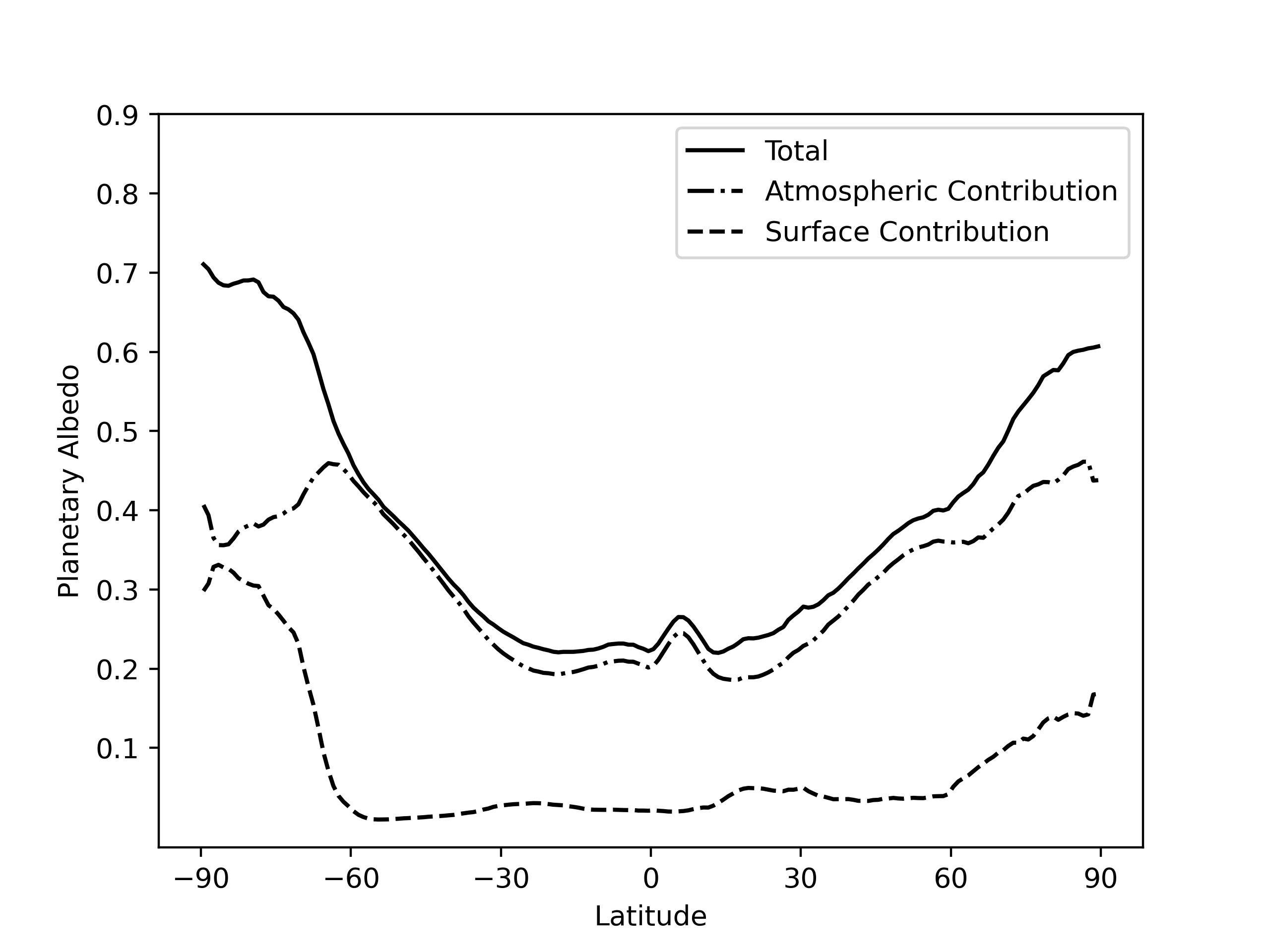}
\caption[Zonal Mean Planetary Albedo]{Zonal mean planetary albedo partitioned between atmospheric and surface components based on CERES EBAF 4.0 data from 3/2000 to 6/2020 c.f. Donohoe and Battisti \cite{donohoe}}
\label{fig:zonalap0}
\end{figure}

 Our modeling of the atmospheric albedo $\alpha_a$ and the ground albedo $\alpha_g$ begins with an initial approximate estimate of the planetary albedo.  As a reference frame and a guide, Figure \ref{fig:zonalap0} shows the zonal mean planetary albedo partitioned between atmospheric and surface components. 

In our model, we first approximate the total planetary albedo by choosing a reference planetary albedo $\alpha_{p0}$ of the form
 \begin{equation}\label{genalbedo}
\alpha_{p0} = \alpha_{p0}^0 + \alpha_{p0}^1x^{4}.
\end{equation}

\begin{figure}[H]
\center
\includegraphics[width=0.8\textwidth]{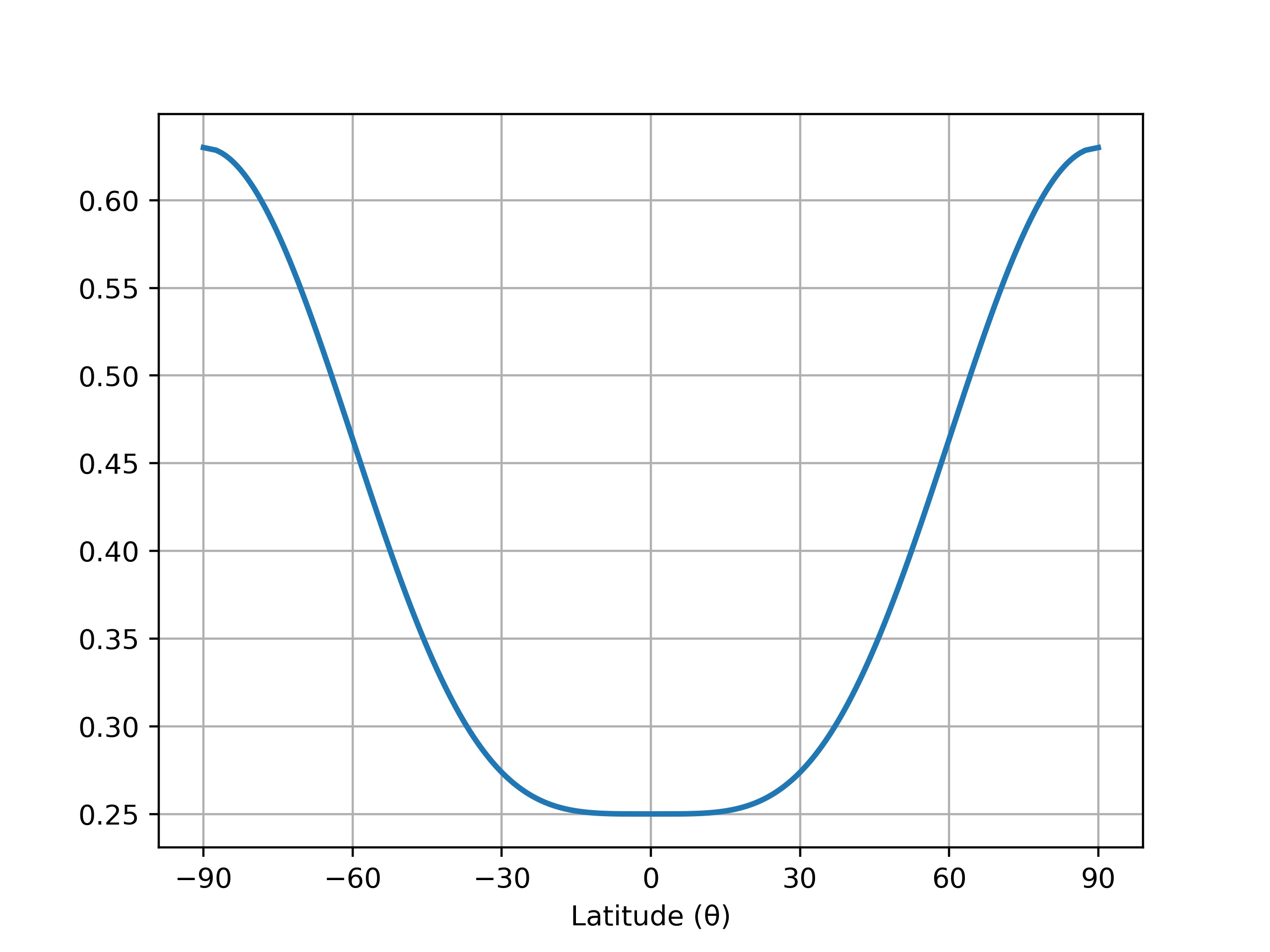}
\caption[Initial Planetary Albedo]{Initial planetary albedo with parameters chosen so that $\alpha_{p0}= 0.25 + 0.38x^{4}.$}
\label{fig:Ap0}
\end{figure}

The coefficients $\alpha_{p0}^0$ and $\alpha_{p0}^1$ are chosen along with parameters for the ground albedo in Eq \eqref{alphag2} so that the average planetary albedo approximates Earth's average planetary albedo, and in order to specify initial equilibrium locations of maximal absolute values of the temperature gradient (for further elaboration, see the third paragraph in Section \ref{results}).
Figure \ref{fig:Ap0} shows a plot of $\alpha_{p0}$ for this choice of parameters:  $\alpha_{p0}^0= 0.25$ and $\alpha_{p0}^1= 0.38$.
 
\begin{figure}[H]
\center
\includegraphics[width=0.7\textwidth]{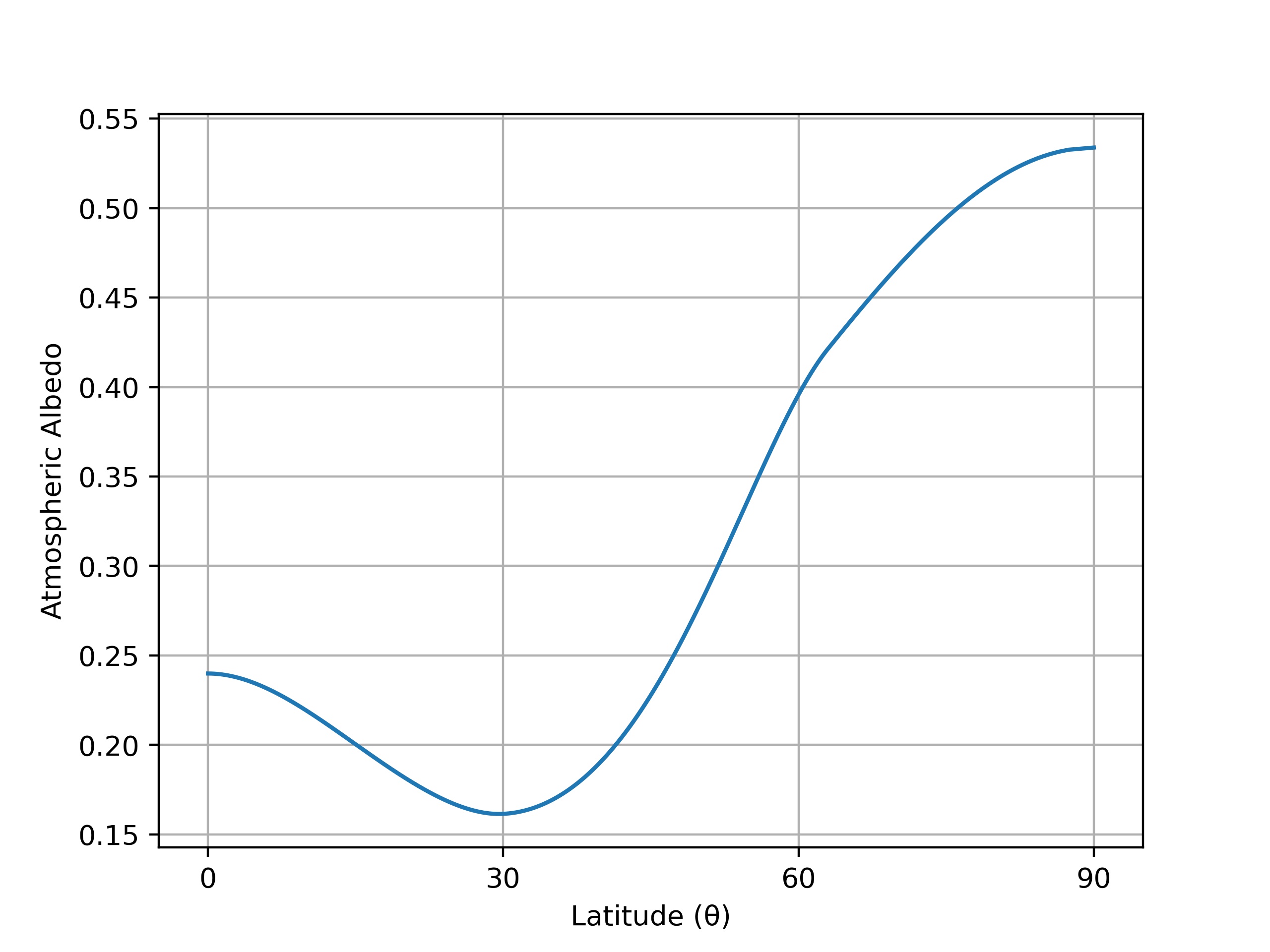}
\caption[Atmospheric Albedo For Cubic Spline $\mathcal{C}_{f_{S3}}$]{Initial atmospheric albedo $\alpha_a$ plotted as a function of latitude from equator to pole for cubic Hermite spline cloud factor $\mathcal{C}_f(\theta)$ using $\alpha_{p0} = 0.25 + 0.38x^{4}$ and the graph in Figure \ref{fig:cfs3}.}
\label{fig:Aaspline}
\end{figure}

We emphasize that at no time step in our computational scheme does the function in Eq. \eqref{genalbedo} represent the planetary albedo in our model, which instead will vary in time in a way that depends on the global temperature distribution. We use $\alpha_{p0}$, along with the cloud factor function $\mathcal{C}_f$, to define the atmospheric contribution to the planetary albedo as:

\begin{equation}
\alpha_{a} = \mathcal{C}_f(\alpha_{p0} - \alpha_{clear})+\alpha_{clear},
\label{eq:atmalb}
\end{equation}

where $\alpha_{clear}$ is the clear sky (cloud free) albedo of the atmosphere which we take as constant, $\alpha_{clear} = 0.149$ \cite{stephens}. An initial sample plot of the atmospheric albedo is given in Figure  \ref{fig:Aaspline}. We note that $\alpha_a$ depends on $\mathcal{C}_f$ and, in turn, $\mathcal{C}_f$ depends on temperature gradients, so $\alpha_a$ depends on temperature gradients.

 We can now define the atmospheric transmittance of short wave radiation (SWR) in terms of $\alpha_a$ as,

\begin{equation}
\mathcal{T}_{sw} = 1 - \alpha_a - A_{sw},
\end{equation}
where $A_{sw} = 0.05$ is the atmospheric absorption of SWR \cite{jentsch}. We note that $\mathcal{T}_{sw}$ depends on $\mathcal{C}_f$, making it interactive.\\

Following other researchers (for example \cite{kaper}), we model the ground albedo as a function of the surface temperature using the hyperbolic tangent function as follows,
\begin{equation}\label{alphag2}
\alpha_{g} = 0.40 - 0.34\tanh(T_s + 8).
\end{equation}

\subsection {Albedo Constraint} \label{albedoconstraint}

 The fraction of incoming solar energy sent back to space from Earth is about $29 \%$ \cite{stephens} with roughly $88\%$ of that coming from the atmospheric contribution and the remainder due to the modulated surface albedo \cite{donohoe, qu}. We therefore tune our model so that our initial atmospheric and modulated ground albedos are close to these values.  They cannot be  constrained in model runs because the atmospheric and ground albedo contributions in our model are dynamic and therefore fluctuate.\\  

 The total planetary albedo $\bar{\alpha}_p$ is given by,

\begin{equation}
\bar{\alpha}_p = \dfrac{1}{2}\int_{-1}^{1}\alpha_p(x)\text{s}(x)dx
\label{eq:ap}
\end{equation}
where, as before, $x$ is the sine of latitude, $\alpha_p(x)$ is the zonally averaged albedo at $x$ given by Eq. \eqref{eq:sumplanalbedo}, and $s(x)$ is the annual weight function for incoming solar radiation given by equation \ref{eq:sx}. The planetary atmospheric albedo $\bar{\alpha}_a$ is defined as, 

\begin{equation}
\bar{\alpha}_a = \dfrac{1}{2}\int_{-1}^{1}\alpha_{a}(x)\text{s}(x)dx,
\label{eq:aa}
\end{equation}
where $\alpha_a(x)$ is the zonally averaged atmospheric albedo at $x$. Therefore, we define the total planetary effective ground albedo by,

\begin{equation}
\overline{\mathcal{T}_{e}\alpha}_g = \bar{\alpha}_p - \bar{\alpha}_a.
\end{equation}

\subsection{The Model}\label{model}
Our model is based on the energy balance equations given in this section.  We begin by linearizing the terms $F_{up}, F_{out}$ in Eqs \eqref{eq:ebm2dima} and \eqref{eq:ebm2dimb} and write,
 \begin{equation}\label{Fs}
\begin{aligned}
F_{out} =& A_{out}+B_{out}T_{a}\\
F_{up} =& A_{up} + B_{up}(T_s - T_a).
\end{aligned}
\end{equation}
Collecting the remaining terms from the preceding sections, the system of coupled PDEs for the zonally and column averaged two layer climate system becomes,

\begin{subequations}
\small
\begin{equation}\label{eq:odeatm}
\begin{aligned}
 C_{a}\dfrac{\partial T_{a}}{\partial t} &=  (1-\alpha_{a}-\mathcal{T}_{sw})\left(1+\dfrac{\alpha_{g}\mathcal{T}_{sw}}{1-\alpha_{a}\alpha_{g}}\right)\dfrac{S_{o}s(x)}{4}+A_{up}+B_{up}(T_{s}-T_{a}) \\
& -A_{out}-B_{out}T_{a} +\dfrac{D_{a}}{a^2} \dfrac{\partial}{\partial x}\left[(1-x^2)\dfrac{\partial T_a{}}{\partial x}\right]
\end{aligned}
\end{equation}
\\
\begin{equation}\label{eq:odesurface}
\begin{aligned}
C_{s}\dfrac{\partial T_{s}}{\partial t}  &= \dfrac{(1-\alpha_{g})\mathcal{T}_{sw}}{1-\alpha_{a}\alpha_{g}}\dfrac{S_{o}s(x)}{4}-A_{up}-B_{up}(T_{s}-T_{a}) \\
&+\dfrac{D_{s}}{a^2} \dfrac{\partial}{\partial x}\left[(1-x^2)\dfrac{\partial T_s{}}{\partial x}\right] 
\end{aligned}
\end{equation}
\\
\begin{equation}
\sqrt{1 - x^{2}}\dfrac{\partial T_{a}}{\partial x}\vline_{x = -1,0,1} = \sqrt{1 - x^{2}}\dfrac{\partial T_{s}}{\partial x} \vline_{x =-1,0,1}= 0 ; t > 0.\label{eq:odeboundary}
\end{equation}
\label{eq:ibvp}
\end{subequations}

Table \ref{tab:par} lists the parameter values for the constants in Eqs. \eqref{eq:odeatm} and \eqref{eq:odesurface}. These are the same values in Rose and Marshall \cite{rose, rose1}, except that our ocean heat capacity is greater by a factor of 10 in order to simulate a greater ocean depth.  The value of $C_s$ in Table \ref{tab:par} together with the nominal values of heat capacity and density for water (as opposed to seawater) of 4184 J/kg/deg C and 1000 kg/m$^3$ assigns an ocean depth of approximately 24 meters. This value is shallow compared to observations of Earth's mixed layer depth \cite{boyer}, but the absence of land in our aqua-planet model is a compensating feature. At any rate, the qualitative behavior of our model is largely independent of the numerical value chosen for $C_s$.
\begin{table}[H]
\begin{center}
\begin{tabular}{||c c c||}
\hline \hline
\textbf{Parameter} & \textbf{Units} & \textbf{Numerical Value} \\ 
a & m & $6.373\times10^{6}$ \\
$S_{0}$ & $W \, m^{-2}$ & 1367 \\
$s_{2}$ & & -0.48 \\
$C_{a}$ & $J \, m^{-2} \, ^{\circ}C^{-1}$ & $10^{7}$\\
$C_{s}$ & $J \, m^{-2} \, ^{\circ}C^{-1}$ & $10^{8}$\\
$D_{a}$ & $W \, ^{\circ}C^{-1}$& $2.7\times10^{13}$ \\
$D_{s}$ & $W \, ^{\circ}C^{-1}$& $5.2\times10^{12}$ \\ 
$B_{up}$ & $W \, m^{-2} \, ^{\circ}C^{-1}$ & 15 \\
$A_{up}$ & $W \, m^{-2}$ & 238 \\
$B_{out}$ & $W \, m^{-2} \, ^{\circ}C^{-1}$ &  1.7 \\
$A_{out}$ & $W \, m^{-2}$ &  variable \\
\hline \hline 
\end{tabular}
\end{center}
\caption[Parameter Values]{Parameter values for the EBM. }
\label{tab:par}
\end{table}

The initial ($t=0$) temperature profile is specified below, and the dynamic feature of the cloud factor function are explained in the next section.\\  

The system of equations Eq.\eqref{eq:odeatm} and \eqref{eq:odesurface}  is defined for  $-1 \leq x \leq 1$, where $x < 0$ is the Southern Hemisphere  and $x > 0$ is the Northern Hemisphere. But since the Southern and Northern hemispheres are symmetric (including our initial conditions), we need only consider the solution from $0 \leq x \leq 1$.

\subsection{Cloud Function Dynamics and Polar Jet Stream}\label{subsectjet}

The response of the eddy-driven jet to arctic amplification \cite{Wang} and changing meridional temperature gradients has been analyzed extensively (e.g., \cite{Barnes, Armour, Francis, Manney, Mbengue,  siler, yin} and references therein). With the thermal wind equations in mind, we identify the mean latitudinal position of the jet stream, at any time $t$, with the location of the  maximum value of a meridional temperature gradient given by,  

\begin{equation}\label{model2grad}
\frac{1}{2}\dfrac{\partial }{\partial \theta} (T_a (t, \theta) + T_s (t, \theta)).
\end{equation}

We motivate this choice as follows.  Let $T(z)=T(\theta, z)$ be the zonally averaged temperature at altitude $z$ and fixed latitude $\theta$.  The vertically averaged temperature $\overline{T}=\overline{T}(\theta)$ at $\theta$ is given by,
\begin{equation}
\overline{T}=\dfrac{1}{h}\int_0^h T(z) dz = \dfrac{1}{h}\int_0^h (T(0) - \Gamma z)\, dz ,
\end{equation}
where $h$ is the height of the troposphere, and the constant $\Gamma$ is the zonally averaged lapse rate at $\theta$.  Thus,
\begin{equation}\label{eval}
\overline{T}=\dfrac{T(0) + (T(0)-\Gamma h)}{2}.
\end{equation}
If we interpret $T_s = T(0)$ and $T(0) - \Gamma h$, to be the atmospheric temperature at altitude $h$ then from Eq.\eqref{eval},
\begin{equation}
\overline{T}=\dfrac{T_s + (T_a - \Gamma h) }{2}.
\end{equation}
The gradient of $\overline{T}$ is then given by Eq.\eqref{model2grad}.\\


We couple the temperature gradient \eqref{model2grad} with the cloud factor function in the following way. We solve the model equations in Sect. \ref{model} numerically by time-stepping out to equilibrium (or quasi-periodicity). $\mathcal{C}_f$ is updated at every timestep by setting it to 0.8 at the latitude of maximum temperature gradient. This choice, together with the cubic Hermite spline functional form and the specified equatorial, polar, and Hadley cell edge values, uniquely determines $\mathcal{C}_f$ at all latitudes. Among other things, it entails that $\mathcal{C}_f$ = 0.8 at all latitudes poleward of the maximum temperature gradient.\\


 For example, the graph in Figure \ref{fig:cfs3} corresponds to a maximum meridional temperature gradient occurring at  $50^{\circ}$ latitude. Since the atmospheric albedo $\alpha_a$ depends on $\mathcal{C}_f$ (c.f. Eq.\eqref{eq:atmalb}), it is updated at each time step.  Similarly, the ground albedo $\alpha_g$ (which is a function of latitude) is updated at each time step according to the values of the surface temperature $T_s$ in the previous time step (see Eq.\eqref{alphag2}).  Numerical approximation details are described in Appendix \ref{program}. \\

\section{Numerical Results}\label{results}

In this section, we present numerical results from our model in response to increases in radiative forcing, such as from increased greenhouse gas concentrations.  Following \cite{rose}, to simulate this, we decrease the parameter $A_{out}$ which controls the flux of outgoing longwave radiation (OLR) from the  top of the atmosphere.
   Our focus is on how the latitudinal locations of the maximum modulus of temperature gradient are affected by  these increases in radiative forcing.  We interpret those latitudes as the averaged locations of the polar jet stream.\\  

Since the coupled partial differential equations of the model are non autonomous, equilibrium temperature and temperature gradient values for each experiment must be found by numerically running them out to equilibrium \footnote{For low values of $A_{\text out}$,  our model does not reach equilibrium with a constant location of the temperature gradient. Instead the maximum temperature gradient becomes quasi-periodic, oscillating between different latitudes, as elaborated below.}.  The results of this section take as initial temperature distributions the final equilibrium temperatures obtained by Rose and Marshall \cite{rose} (in their Figure 2), but the model behaviors are insensitive to the choice of initial temperature distributions.\\  

\begin{figure}[H]
\hspace*{-1.5cm}
%
		\subfigure[]{%
			\label{sum214}
			\includegraphics[width=0.6\textwidth]{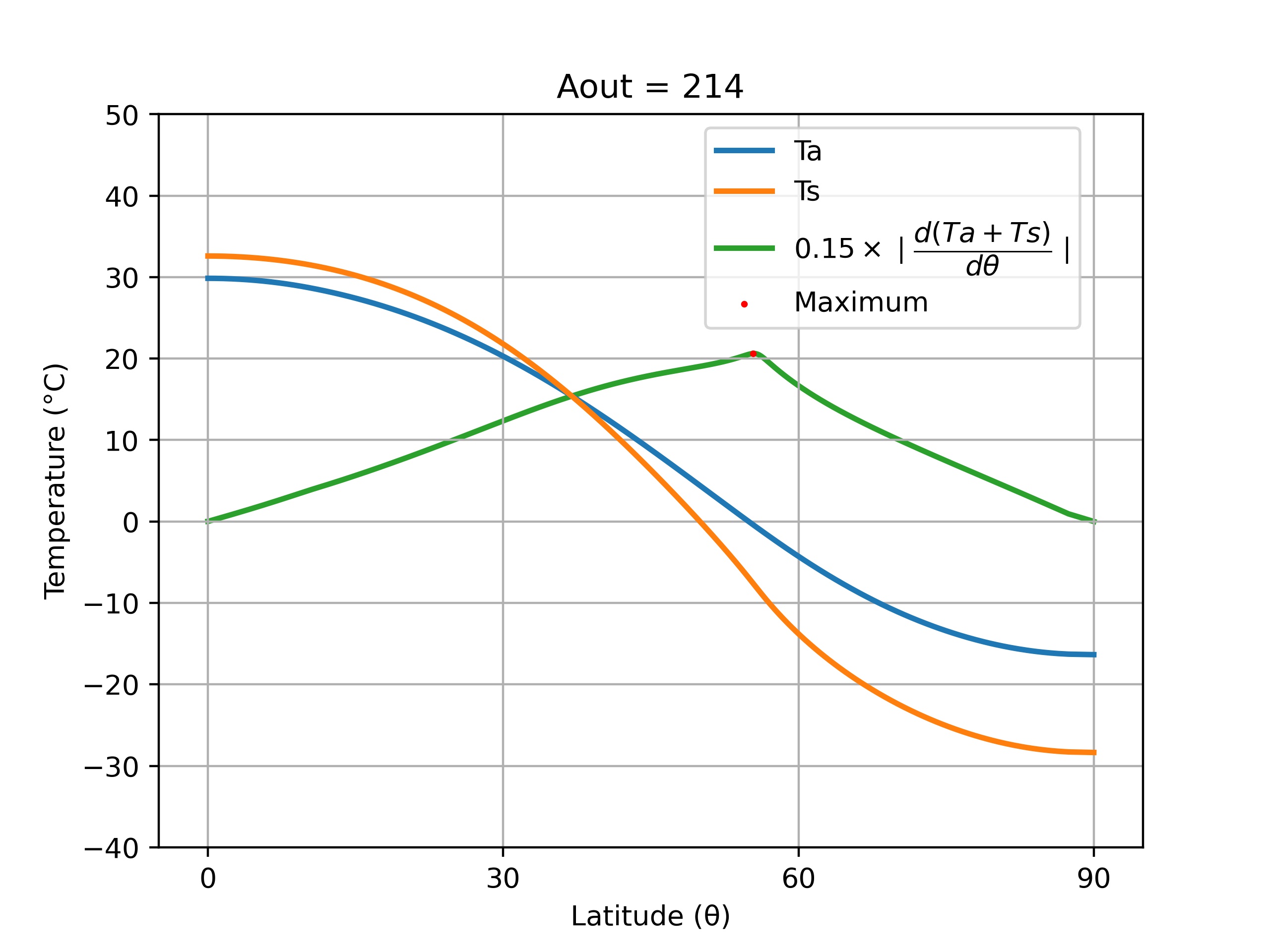}
		}%
		\subfigure[]{%
			\label{sum210}
			\includegraphics[width=0.6\textwidth]{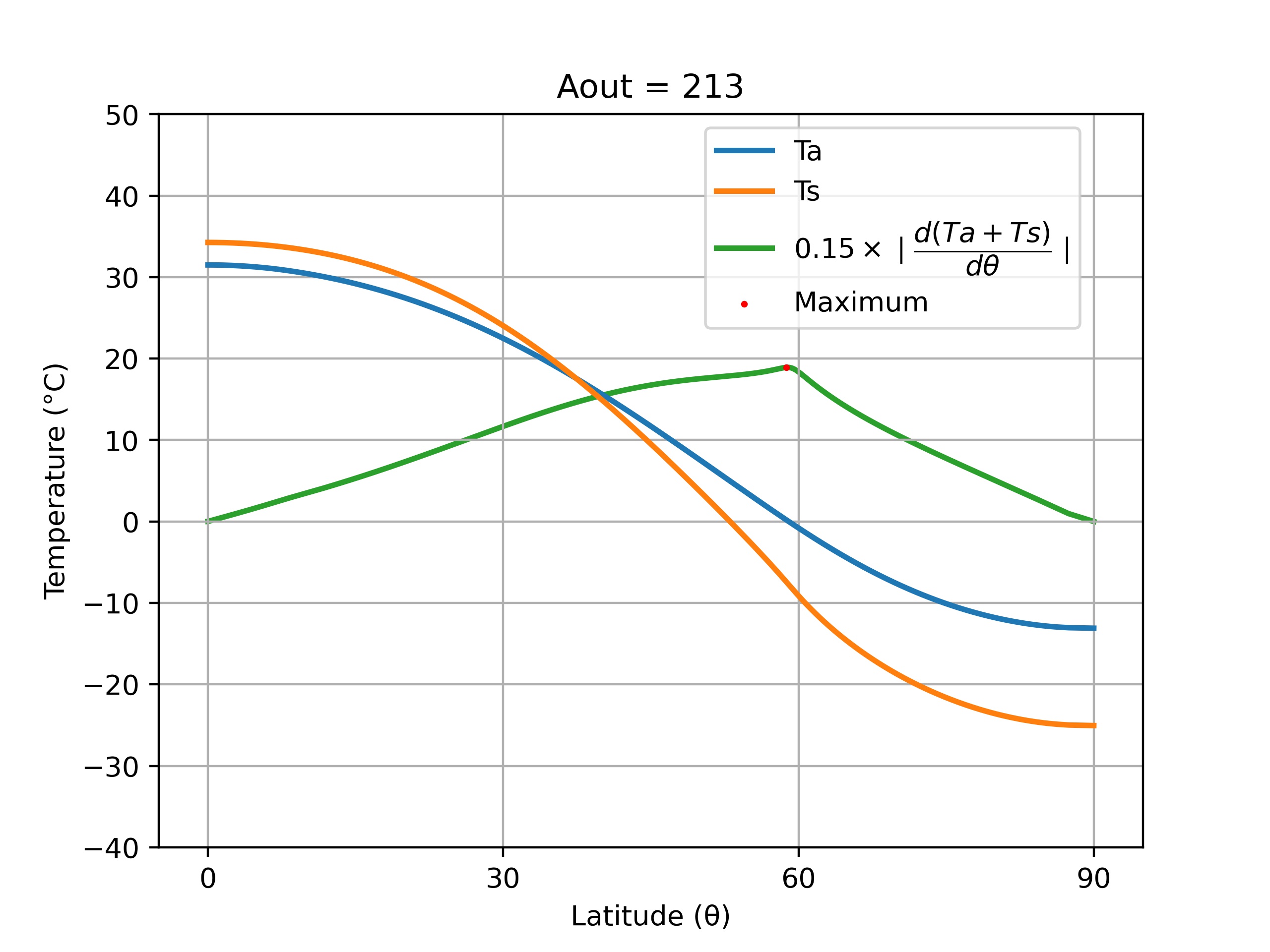}
		}\\
		\hspace*{-1.5cm}
		\vspace{-1\baselineskip}
		\subfigure[]{%
			\label{sum208}
			\includegraphics[width=0.6\textwidth]{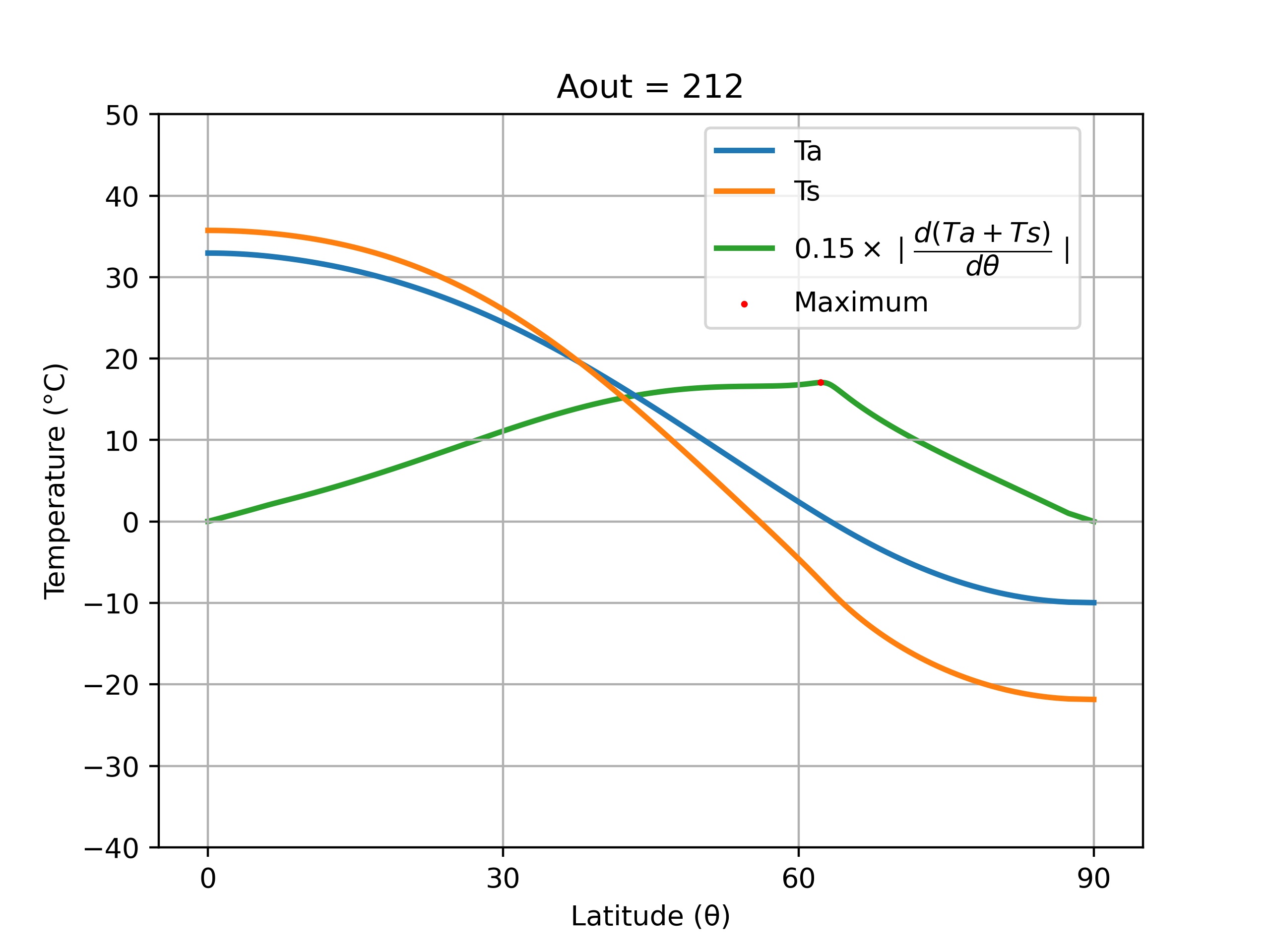}
		}%
		\subfigure[]{%
			\label{sum207}
			\includegraphics[width=0.6\textwidth]{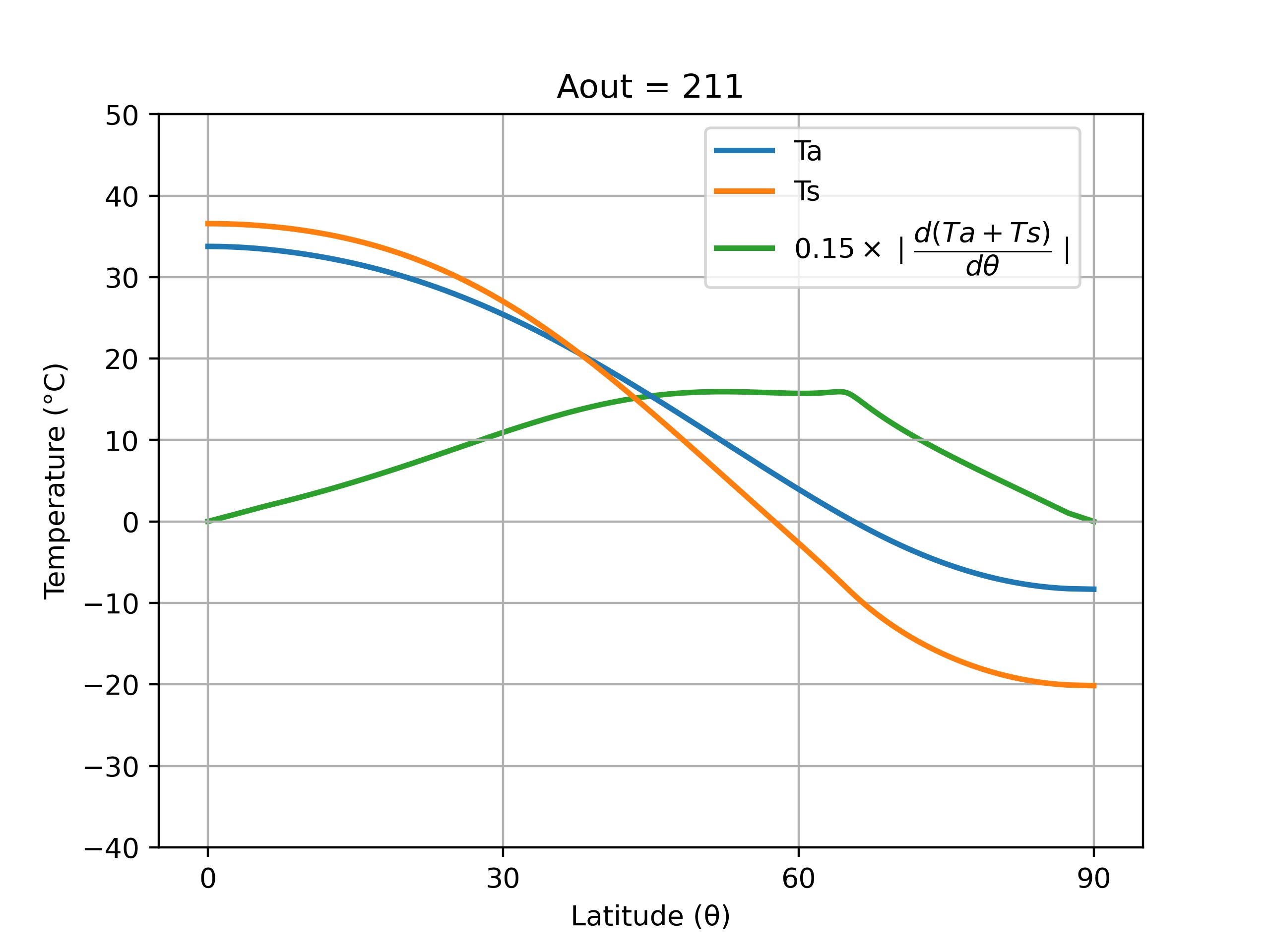}
		}\\
	\caption{%
	Meridional profiles of $T_s$ and $T_a$ with temperature gradients.  Increased greenhouse gas concentrations are modeled by decreasing values of $A_{out}$. The maximum value of $|\partial/\partial \theta (T_a + T_s)|$ (scaled by a factor of $0.15$ for display purposes)  represents the average latitude of the polar jet under the indicated forcings of $A_{out}$. In plots (a) through (c), the jet moves poleward monotonically as $A_{out}$ decreases, but the gradient in plot (d) begins to form an approximate plateau with oscillatory equilibrium location. The graph shown in subfigure (d) is for a time step at which the maximum value of $|\partial/\partial \theta (T_a + T_s)|$ occurs at latitude 64.027.}%
	\label{sumfigures}
\end{figure}

To set a reference climate, we take $A_{out}$ = 214 Wm$^{-2}$.  In equilibrium, this results in a climate with a planetary albedo, $\bar{\alpha}_p = 0.298$, and average temperatures given by $T_s = 14.4^\circ$C and $T_a = 15.5^\circ$C.  The maximum absolute value of the atmospheric temperature gradient occurs at $55.4^\circ$ latitude.  This is our proxy for the average latitude of the jet stream. The temperature and gradient distributions for $A_{out}$ = 214 Wm$^{-2}$ are displayed in Figure \ref{sum214} and Table \ref{tableclimsen}.\\

By decreasing the parameter $A_{out}$, we introduce a longwave radiative forcing in the model. Meridional profiles of $T_s$ and $T_a$ and temperature gradient plots for $A_{out} = 214, 213, 212$, and $211$ Wm$^{-2}$ are shown in Figure \ref{sumfigures}. \\ 

\begin{figure}[H]
\hspace*{-1.5cm}
%
		\subfigure{%
			\label{sum211}
			\includegraphics[width=0.6\textwidth]{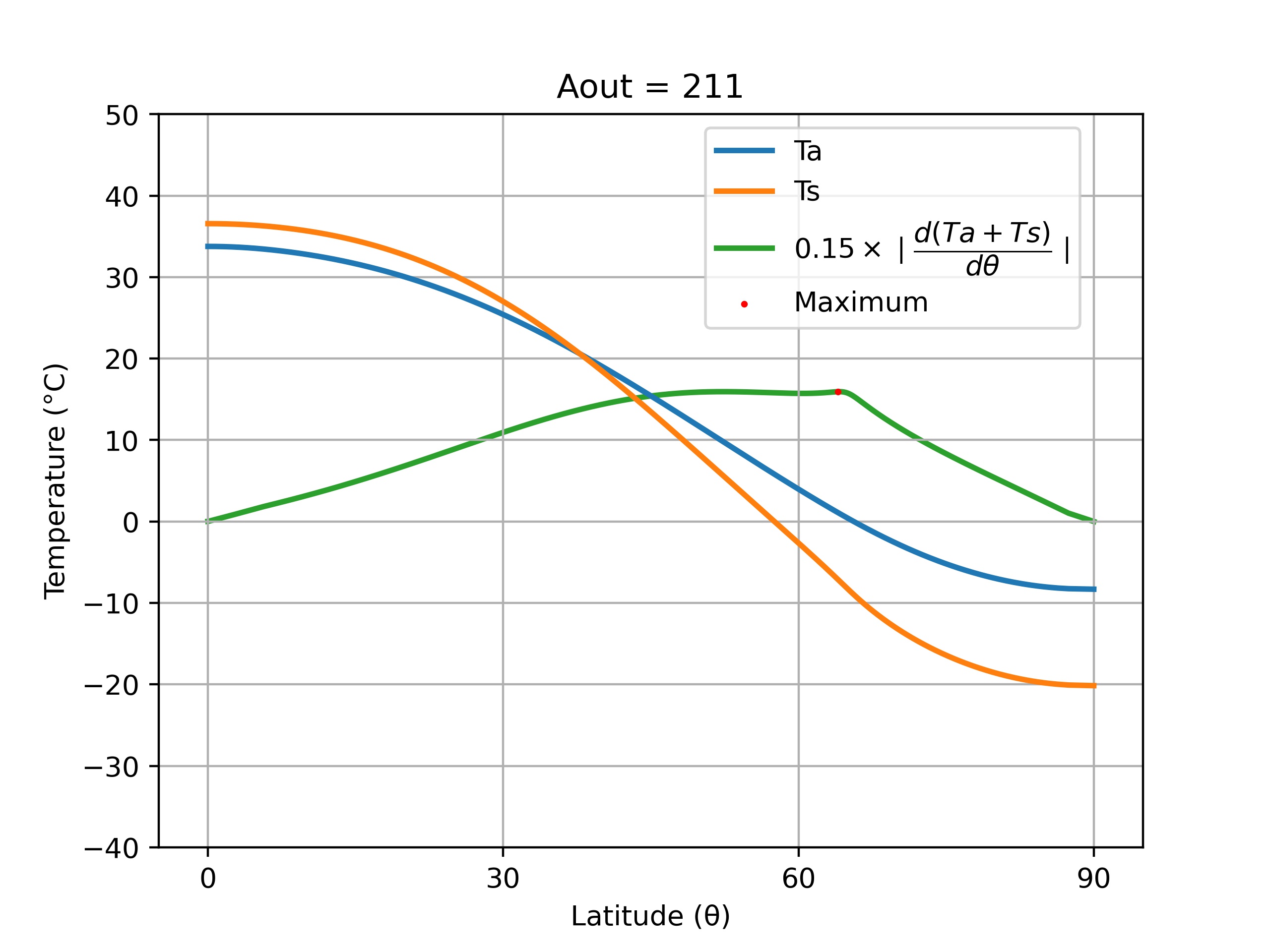}
		}%
		\subfigure{%
			\label{sum210}
			\includegraphics[width=0.6\textwidth]{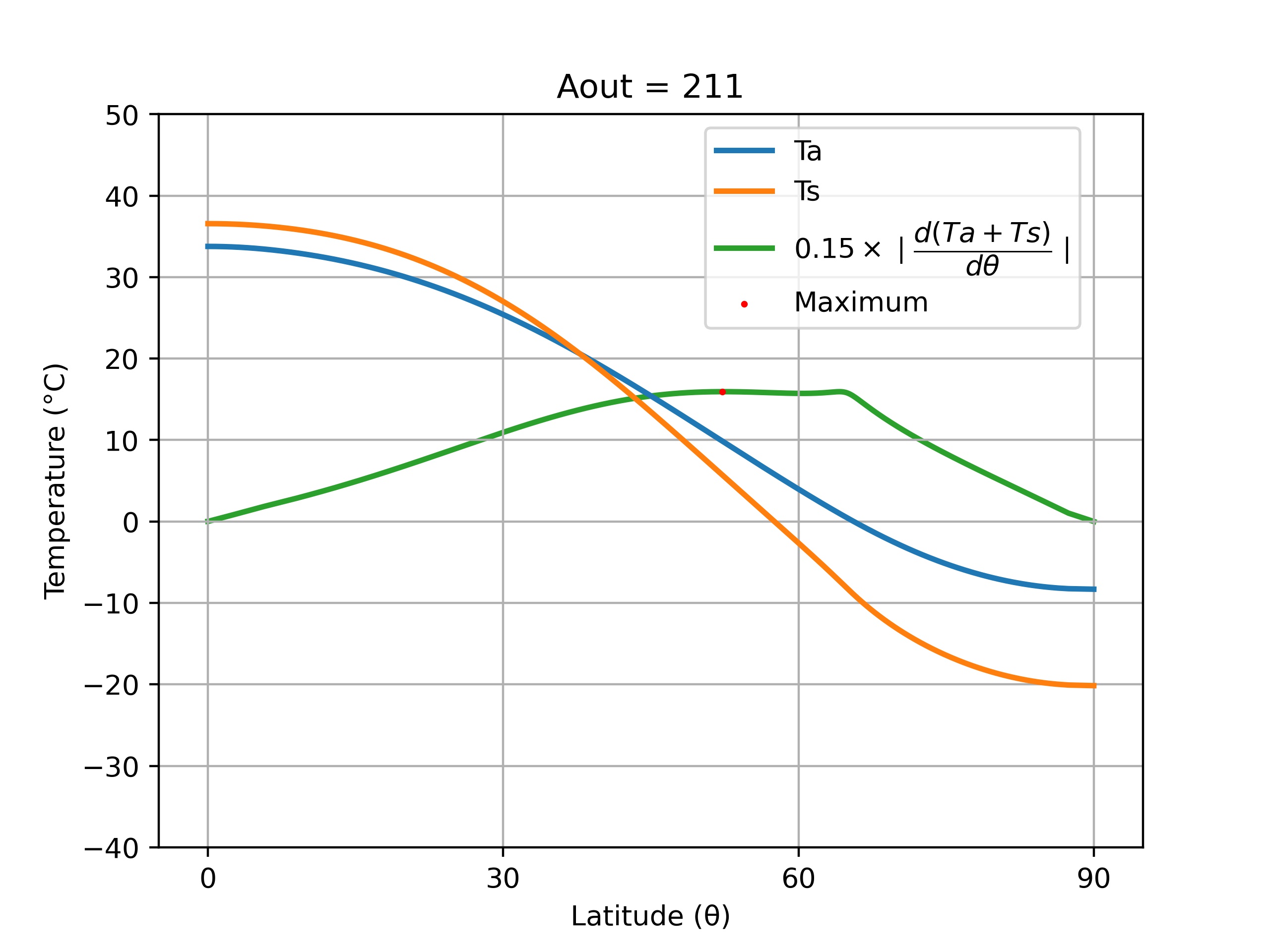}
		}\\
	\caption{%
	Meridional profiles of $T_s$ and $T_a$ with temperature gradients for $A_{out} = 211$ Wm$^{-2}$ at two different time steps beyond 8725 days.  $A_{out} = 211$ Wm$^{-2}$ is the largest integer value of $A_{out}$ at which oscillatory behavior of the maximum value of $|\partial/\partial \theta (T_a + T_s)|$ occurs.}%
	\label{sumfigures2}
\end{figure}

Equilibrium is reached for the first three forcings, $A_{out} = 214, 213, 212$ Wm$^{-2}$. However, for $A_{out} = 211$ Wm$^{-2}$, the maximum absolute value of the atmospheric temperature gradient begins to exhibit oscillatory behavior.  This is indicated by the red dots in Figure \ref{sumfigures2}.  

As  $A_{out}$ continues to decrease to values below 211 Wm$^{-2}$ (so that radiative forcing increases), the modulus of the temperature gradient given by Eq.\eqref{model2grad} does not peak at a singular latitude, but instead produces a collection of nearly equal large values within an interval of latitudinal coordinates.  As a physical interpretation, this suggests oscillatory behavior of the jet stream, and this is shown graphically in Figure  \ref{sumfigures4}, based on the data in Appendix \ref{quasiperiodic}. Additional detail is shown in Figure \ref{sumfigures3}, which displays plots of the temperature gradient within a narrower range of latitudes, and illustrates the formation of approximate plateaus of maximum values of $|\partial (T_a + T_s)/\partial \theta |$. 
We note that numerical experiments show that the same oscillatory behavior appears when the time step is reduced to half days and quarter days (instead of days).\\

Tables \ref{tableclimsen} and \ref{tableclimsen2} show the mean latitudinal locations of the jet, standard deviations from the means of the jet locations, along with temperature and albedo data, as  $A_{out}$ decreases from 214 to 202 Wm$^{-2}$.  The standard deviations reveal the extent of oscillations of the jet.  As shown in Table \ref{tableclimsen}, oscillations increase as $A_{out}$ decreases to 208 Wm$^{-2}$. The movement of the jet location, as the forcing increases, is initially poleward, but as the forcing increases further, the mean jet location begins to move equatorward.  

\begin{table}[htp]
\caption{Model Data for Low Forcings and Increasing Jet Oscillations}  
\begin{center}
\resizebox{\textwidth}{!}{
\begin{tabular}{||c|c |c |c |c|c|c||}
\hline
$A_{\text out}$ in Wm$^{-2}$ & $214$ & $213$ & $212$ &$211$&$210$&$209$\\
\hline
Mean Jet Latitude &55.4$^\circ$ & 58.8$^\circ$ & 62.3$^\circ$ & 61.9$^\circ$ & 58.3$^\circ$ & 54.8$^\circ$ \\
Standard Deviation &0$^\circ$&0$^\circ$&0$^\circ$&4.51$^\circ$&6.64$^\circ$&7.39$^\circ$\\
Planetary albedo  & 0.30 & 0.29 & 0.28 & 0.28 & 0.28 & 0.28\\
Global Ave $T_s$ &14.4$^\circ$C & 17.0$^\circ$C & 19.3$^\circ$C & 20.5$^\circ$C & 20.6$^\circ$C & 20.6$^\circ$C\\
Global Ave $T_a$ & 15.5$^\circ$C & 17.9$^\circ$C & 20.0$^\circ$C & 21.1$^\circ$C & 21.2$^\circ$C & 21.3$^\circ$C \\
\hline
\end{tabular}
}
\end{center}\label{tableclimsen}
\end{table}%

\vspace{-.3in}

\begin{table}[htp]
\caption{Model Data for High Forcings and Decreasing Jet Oscillations}  
\begin{center}
\resizebox{\textwidth}{!}{
\begin{tabular}{||c|c|c |c |c|c |c|c||}
\hline
$A_{\text out}$ in Wm$^{-2}$ &$208$& $207$ & $206$ & $205$& $204$ &$203$&$202$\\
\hline
Mean Jet Latitude &51.2$^\circ$ &47.3$^\circ$ & 43.4$^\circ$ & 42.5$^\circ$& 41.6$^\circ$ & 40.9$^\circ$ & 40.2$^\circ$ \\
Standard Deviation &7.22$^\circ$&5.87$^\circ$&0.32$^\circ$&0.31$^\circ$&0.31$^\circ$&0.30$^\circ$&0.30$^\circ$\\
Planetary albedo  & 0.28 & 0.28 & 0.29 & 0.29 & 0.29 & 0.29 & 0.28\\
Global Ave $T_s$ & 20.6$^\circ$C &20.6$^\circ$C & 20.7$^\circ$C & 21.6$^\circ$C & 22.5$^\circ$C  & 23.4$^\circ$C& 24.3$^\circ$C \\
Global Ave $T_a$ & 21.4$^\circ$C & 21.5$^\circ$C & 21.6$^\circ$C & 22.5$^\circ$C & 23.3$^\circ$C& 24.2$^\circ$C & 25.1$^\circ$C \\
\hline
\end{tabular}
}
\end{center}\label{tableclimsen2}
\end{table}%


\begin{figure}[H]
\hspace*{-1.5cm}
%
		\subfigure[]{%
			\label{207}
			\includegraphics[width=0.6\textwidth]{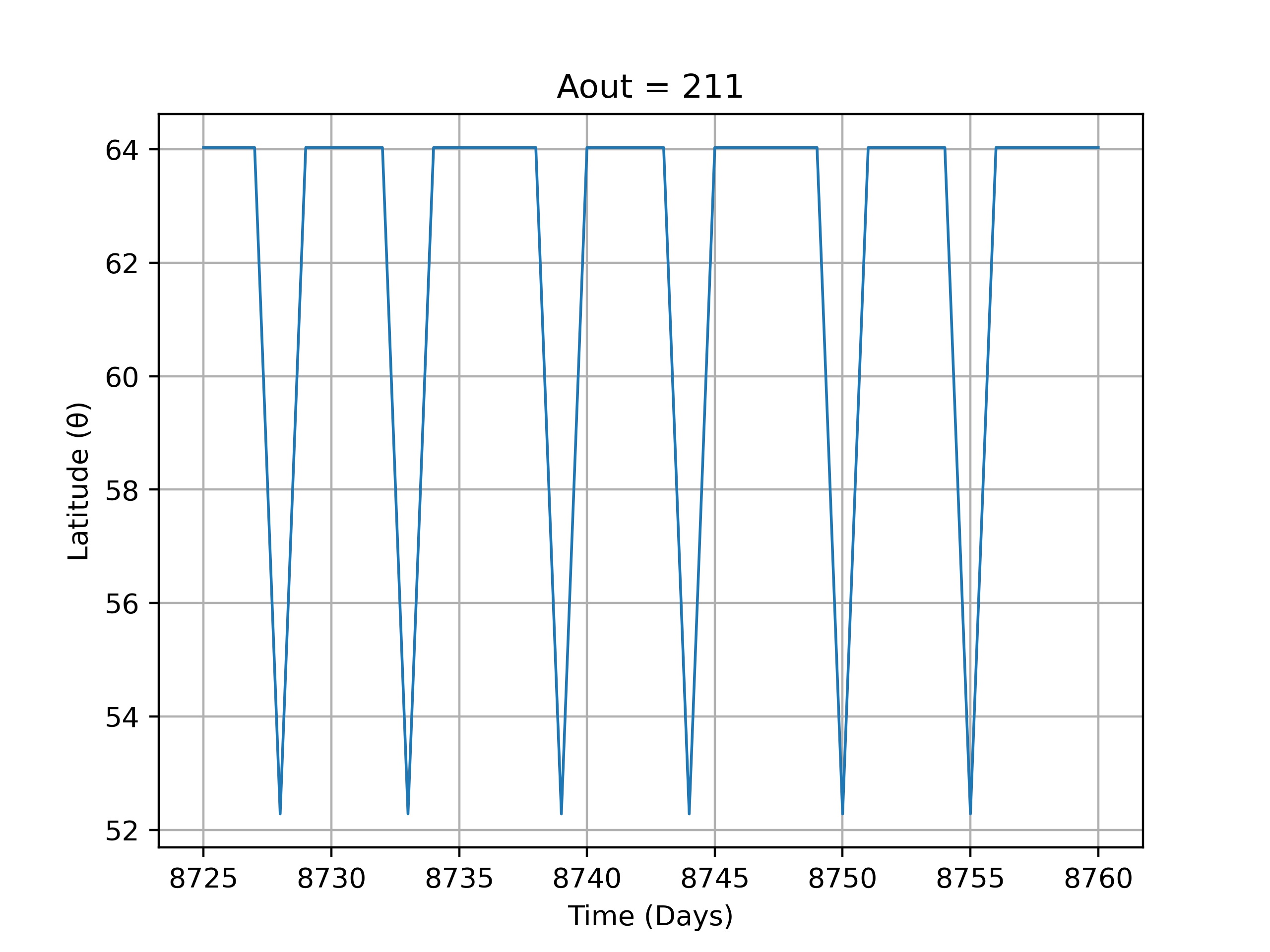}
		}%
		\subfigure[]{%
			\label{206}
			\includegraphics[width=0.6\textwidth]{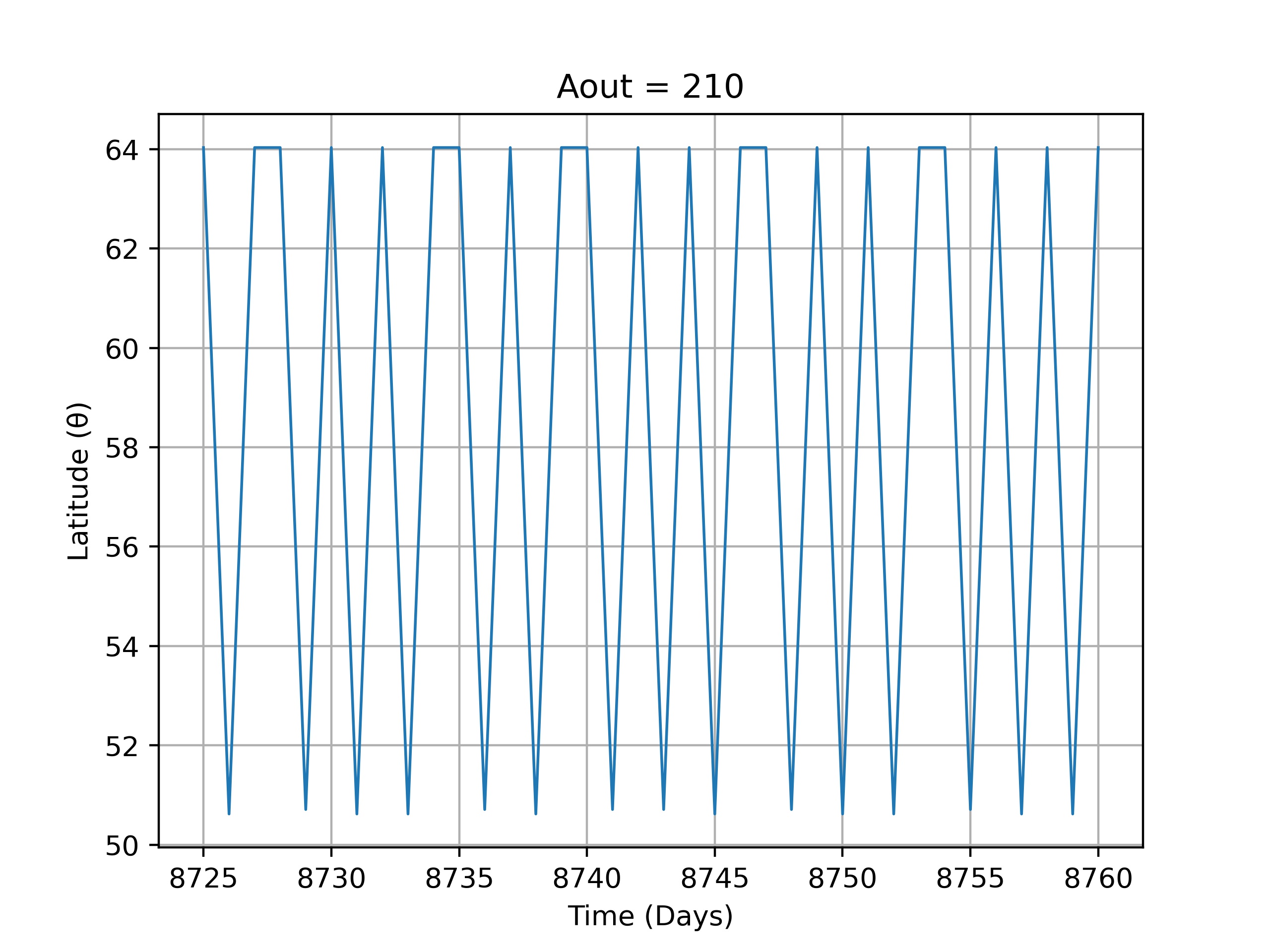}
		}\\
		\hspace*{-1.5cm}
		\vspace{-1\baselineskip}
		\subfigure[]{%
			\label{207}
			\includegraphics[width=0.6\textwidth]{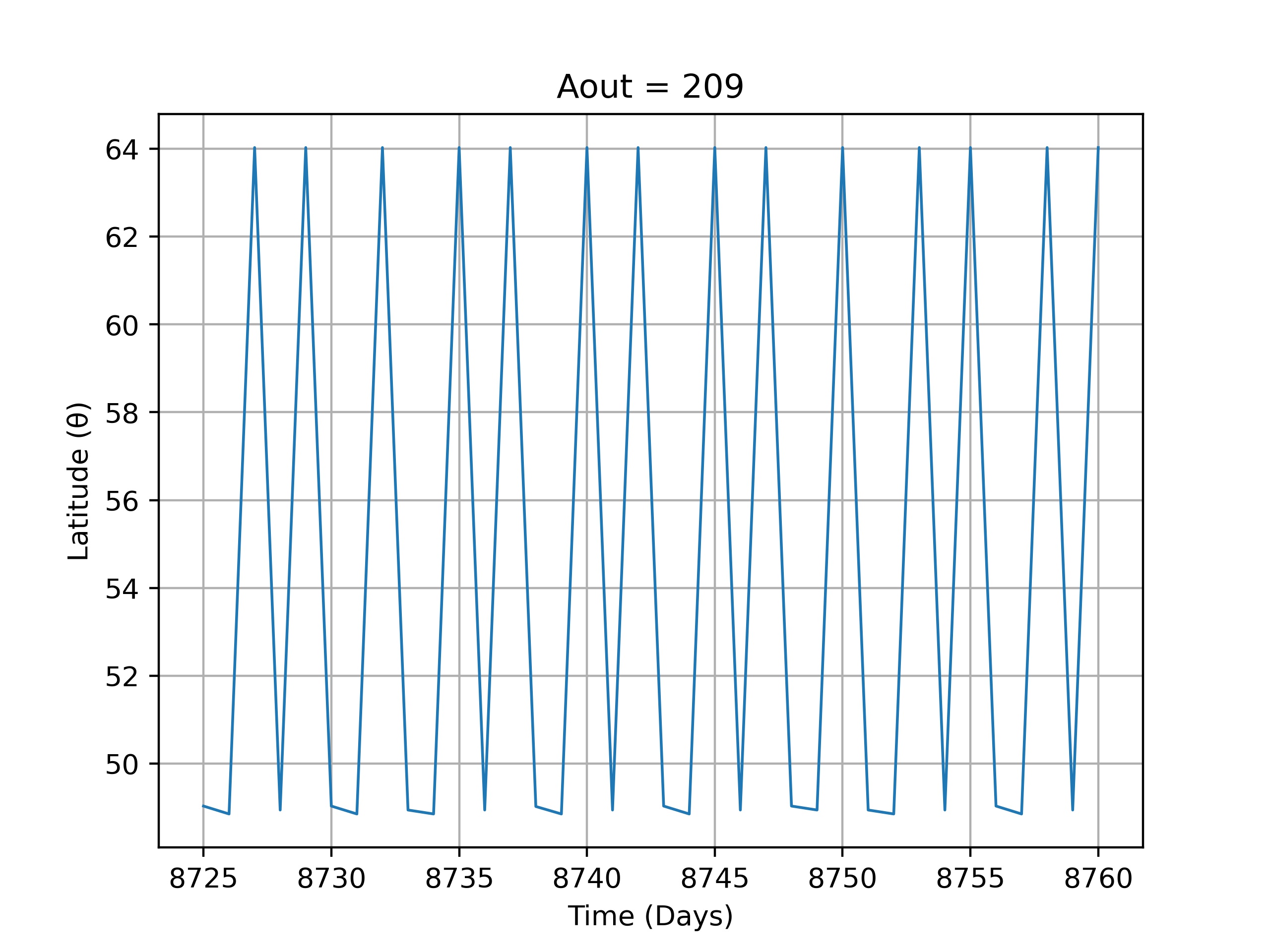}
		}%
		\subfigure[]{%
			\label{sum207}
			\includegraphics[width=0.6\textwidth]{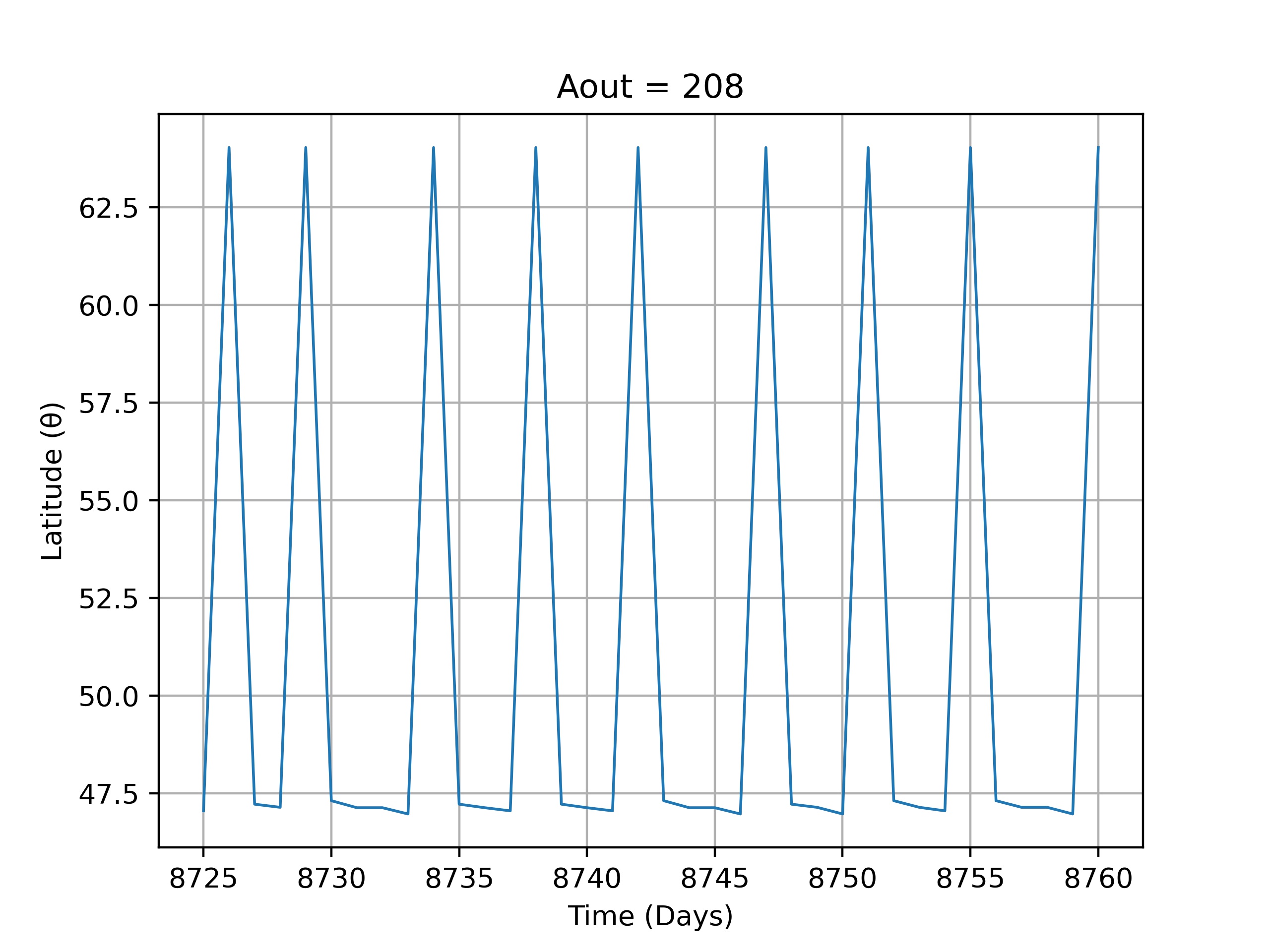}
		}\\
	\caption{%
Quasi-periodic graphs of $\max|\partial/\partial \theta (T_a + T_s)|$ as functions of time (in days) for large forcings corresponding to radiative forcings determined by: (a) $A_{out}= 211  Wm^{-2}$, (b) $A_{out}= 210 Wm^{-2}$,  (c) $A_{out}= 209 Wm^{-2}$,  (d) $A_{out}= 208 Wm^{-2}$.} %
	\label{sumfigures4}
\end{figure}

\begin{figure}[H]
\hspace*{-1.5cm}
%
		\subfigure{%
			\label{sum207z}
			\includegraphics[width=0.6\textwidth]{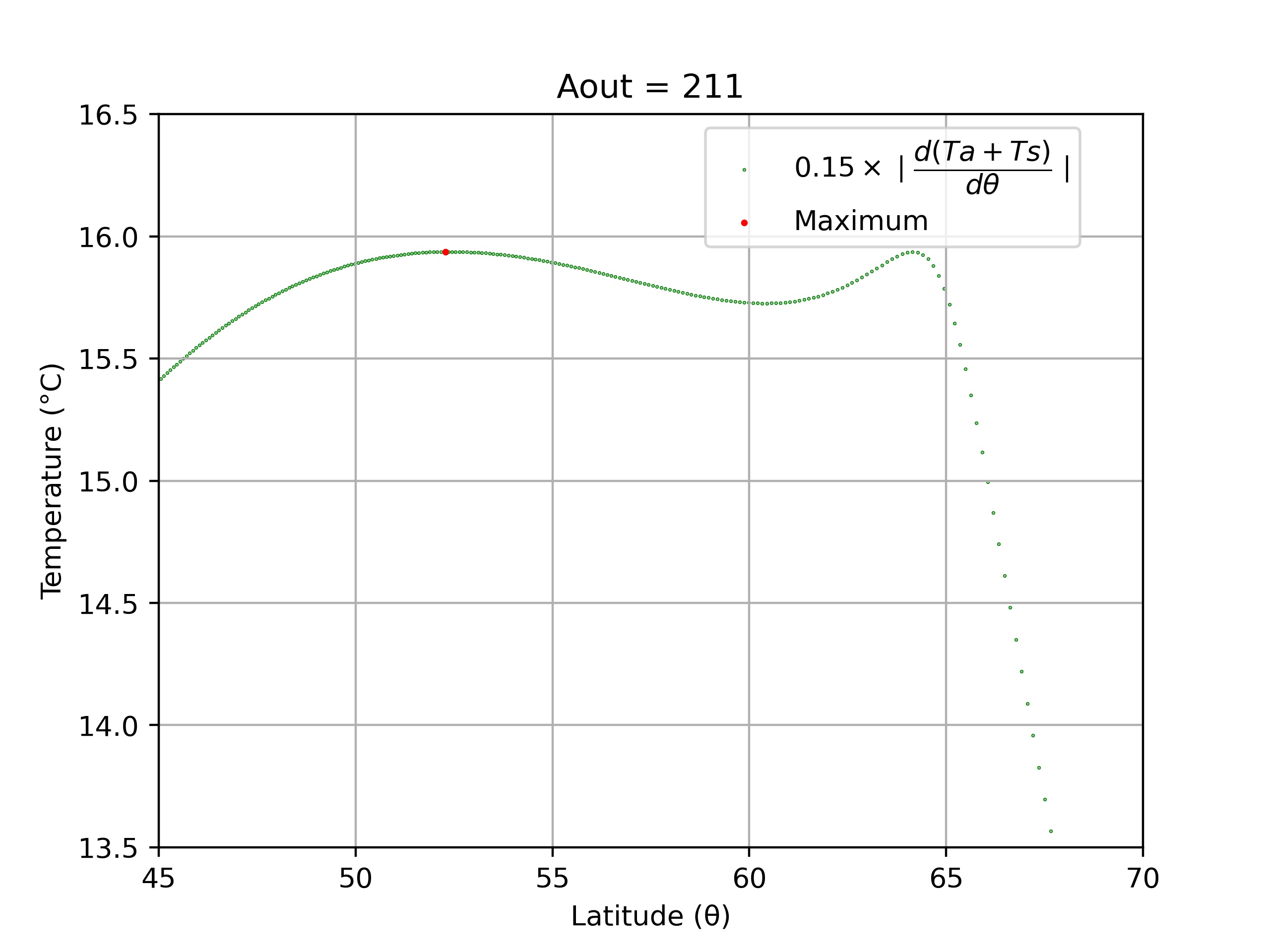}
		}%
		\subfigure{%
			\label{sum206}
			\includegraphics[width=0.6\textwidth]{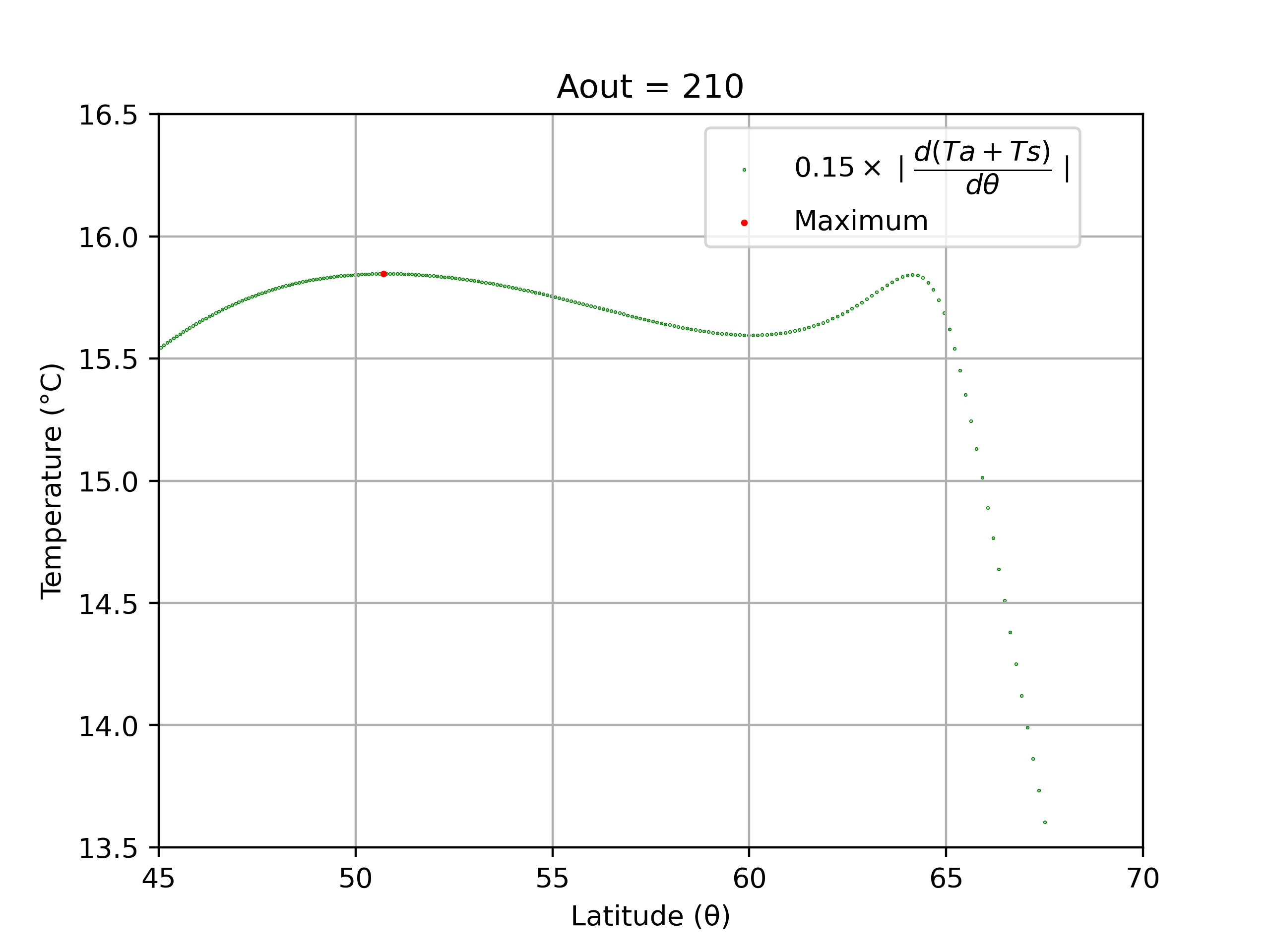}
		}\\
		\hspace*{-1.5cm}
		\vspace{-1\baselineskip}
		\subfigure{%
			\label{sum205}
			\includegraphics[width=0.6\textwidth]{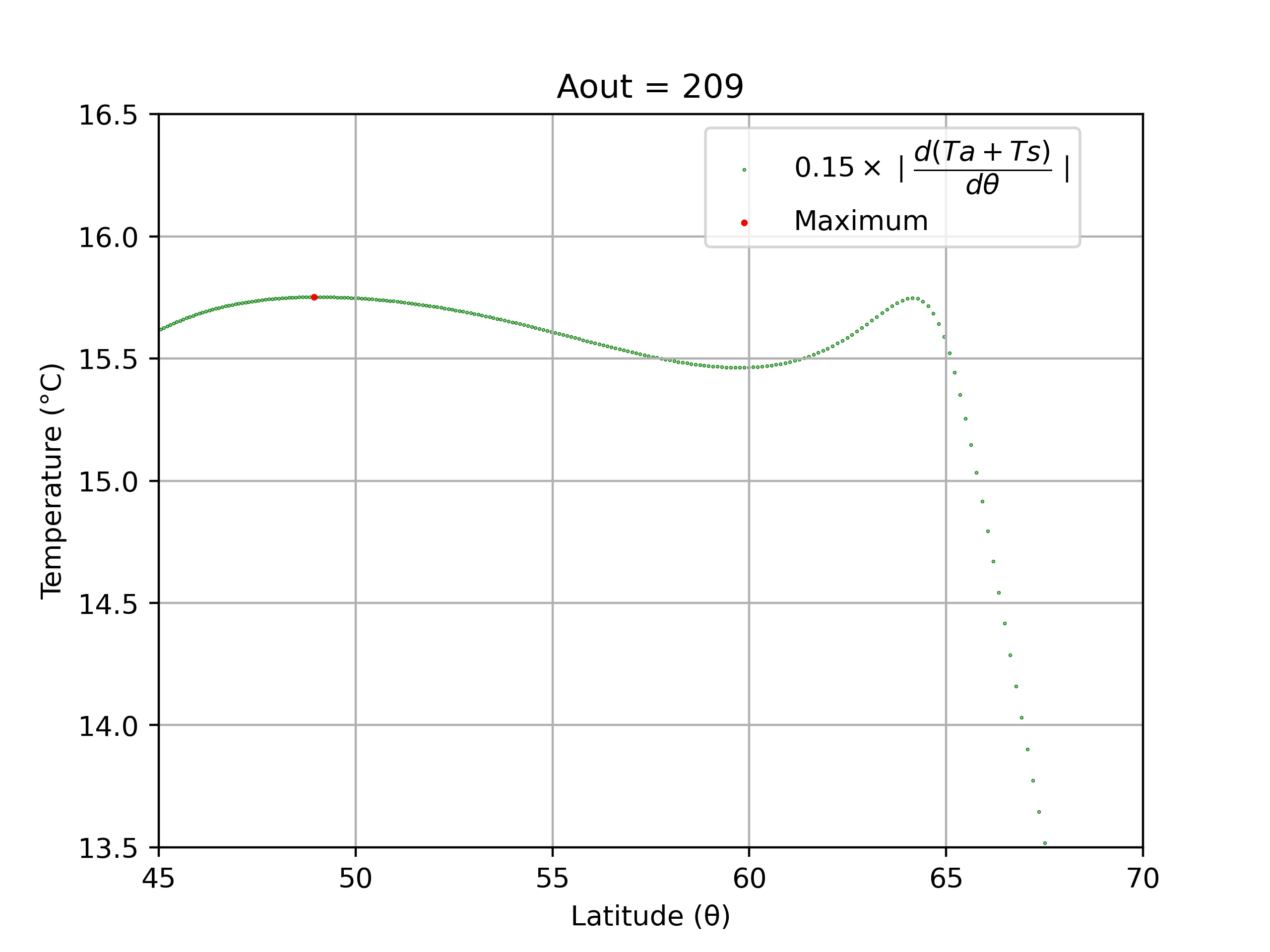}
		}%
		\subfigure{%
			\label{sum204}
			\includegraphics[width=0.6\textwidth]{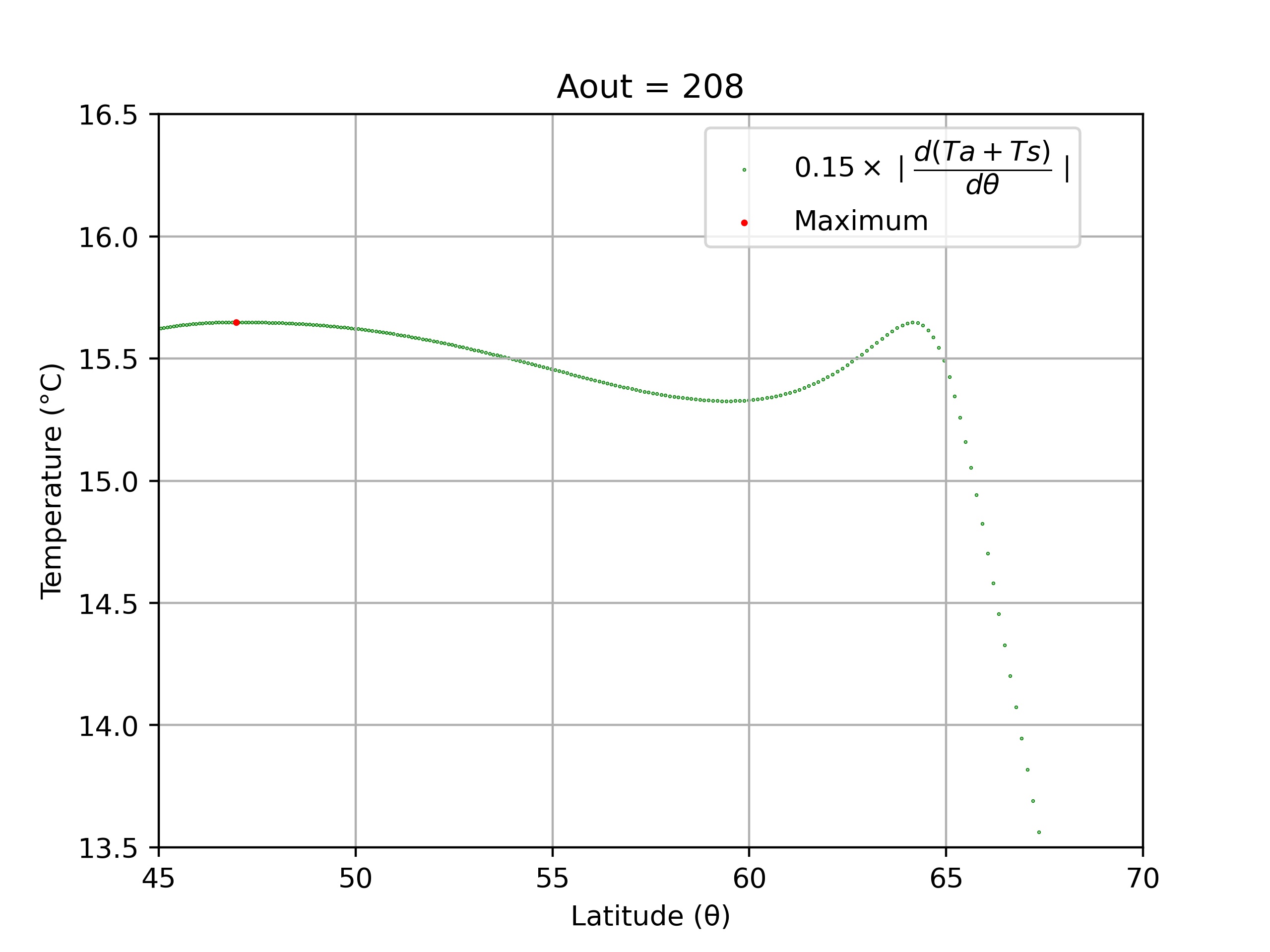}
		}\\
	\caption{%
	Sample maxima of $|\partial (T_a + T_s)/\partial \theta |$ indicated with a red dot, along with displays of approximate plateaus of maximum values, at particular time steps for values of  $A_{out}< 212$ Wm$^{-2}$ for which the jet is oscillatory. With each time step, the maximum on each plot shifts to a different latitude.} 
	\label{sumfigures3}
\end{figure}
Table \ref{tableclimsen2} shows that the mean jet location continues to move equatorward as the forcing increases (i.e., as $A_{\text out}$ decreases), and the standard deviation data indicates that the jet oscillations decrease and nearly cease at $A_{\text out} = 206$ Wm$^{-2}$, and lower values, so that the mean jet location is nearly constant for each of those values.\\  

The picture that emerges is that significant oscillations of the jet occur only for the band of $A_{\text out}$ values between 211 and 207 Wm$^{-2}$, and the mean location of the jet increases poleward from 55.4$^\circ$ latitude for $A_{\text out} =214$ Wm$^{-2}$ to 62.3$^\circ$ latitude for $A_{\text out} =212$ Wm$^{-2}$, and thereafter moves equatorward. \\
 
The climate sensitivity of our model can be determined from the temperature data in Tables \ref{tableclimsen} and \ref{tableclimsen2}.  For the purpose of comparison, we first note that the IPCC's AR6 estimate for Earth's modern Effective Radiative Forcing (ERF) for a doubling of atmospheric CO$_2$ is $3.93 \pm 0.47$ Wm$^{-2}$, and $3.73 \pm 0.44$ Wm$^{-2}$ for the stratospherically adjusted radiative forcing. The equilibrium climate sensitivity is estimated to be $3^\circ$C \cite{foster}.\\ 

Tables \ref{tableclimsen} and \ref{tableclimsen2} show that the climate sensitivity of our model varies with temperature and forcing.  This is not unprecedented.  In their study of climate sensitivity in the context of high temperature and large radiative forcings, Caballero and Huber \cite{caballero} gave evidence that hothouse climate states may have different climate sensitivities per doubling of CO$_2$ than Earth's present state.  In their study of early Paleogene and possible future high temperature modern climates, the temperature gain with each doubling of CO$_2$ was not constant according to their model, but instead increased with increasing CO$_2$ concentrations.\\

By contrast, climate sensitivity of our model varies strongly and nonmonotonically with temperature and $A_{\text out}$, encompassing values that are both above and below reasonable estimates of the modern Earth's climate sensitivity, but also Earthlike sensitivity at high temperatures and forcings.  Unit increases in forcing, from $A_{\text out}= 208$ to 207 and $A_{\text out}= 207$ to 206 Wm$^{-2}$ each result in an increase of the atmospheric temperature $T_a$ by only 0.1$^\circ$C.  But unit increases from lower values of  $A_{\text out}$, corresponding to higher temperatures, shown in Table \ref{tableclimsen2} result in increases of 0.9 and 0.8$^\circ$C (corresponding to climate sensitivities of  3.5$^\circ$C and 3.1 $^\circ$C respectively, assuming the  IPCC's reported effective radiative forcing of 3.9 Wm$^{-2}$ for a doubling of CO$_2$).\\

Data in Table \ref{tableclimsen} reveals an unrealistically high climate sensitivity for the larger consecutive values of $A_{\text out}$ compared to Earth's modern climate, and unrealistically low climate sensitivity for smaller consecutive values of $A_{\text out}$. An increase in forcing from $A_{\text out}= 214$ to 213 Wm $^{-2}$ results in an increase of $T_a$ by 2.4$^\circ$C, but as $A_{\text out}$   decreases, the temperature increases decline. For consecutive large values of $A_{\text out}$, it is likely that positive shortwave feedbacks, created by the interactive clouds and surface albedo, nearly cancel out the negative feedback associated with the increase of outgoing longwave radiation with increasing temperatures. \\

We note, however, that over the full range of $A_{\text out}$ values, the average climate sensitivity is evidently closer to Earth-like climate sensitivity. Comparing the data for $A_{\text out} = 214$ and $A_{\text out} = 202$ Wm$^{-2}$, the ratio of temperature ($T_a$) increase per unit forcing is,
\begin{equation}
\frac{25.1-15.5}{214-202}= 0.8\frac{^\circ \text{C}}{\text{Wm}^{-2}},
\end{equation}
which amounts to a warming of 3.1$^\circ$C from the IPCC's reported effective radiative forcing of 3.9 Wm$^{-2}$ for a doubling of atmospheric CO$_2$.

\section{Discussion}\label{conclude}

Our results may be compared with observations and predictions from more elaborate models.  Using the Coupled Model Intercomparison Project (CMIP5) and assuming the representative concentration pathway 8.5 (RCP8.5) scenario, Barnes and Polvani \cite{Barnes} found that all jets migrate poleward in the twenty-first century.  Using reanalysis, Manney and Hegglin \cite{Manney} found that the southern polar jet has shown a robust poleward shift, while the northern polar jet has shifted equatorward in most regions and seasons. Liu et al. showed in \cite{LiuW} that, in a simulation of the Last Glacial Maximum, NCAR's CCSM4 model indicates that, in the Southern Hemisphere, the ice line  advances equatorward while the jet shifts poleward. In \cite{Francis} Francis and Vavrus found evidence to support a linkage between rapid Arctic warming and more frequent high-amplitude, wavy jet-stream configurations (though they considered zonally asymmetric aspects of the flow which our model does not simulate), and in \cite{karamperidou} Karamperidou, Cioffi, and Lall considered meridional surface temperature gradients and found them to be determinants of large-scale atmospheric circulation patterns. \\

The behavior of our model shares qualitative features with these investigations.  An increase in radiative forcing, as from increased greenhouse gas concentrations, results in an initial poleward movement of the polar jet, followed by a equatorward shift of averaged locations and quasi-periodic oscillations, under greater forcings. Our results may also be compared to those of MS18 \cite{Mbengue} and SAR18 \cite{siler}, both of which used EBMs to demonstrate the influence of changing Hadley cell boundaries on the location of mid-latitude storm tracks.  Our results do not contradict those findings but suggest that the latitudinal distribution of clouds may play a significant role as well. \\ 

More broadly, the cloud factor function in our model may be regarded as a prototype for further investigations.  The cubic Hermite spline used to define the cloud factor function in this article depends only on a small number of fixed values, those at the equator, the Hadley cell boundary, the pole, and at the location of the maximum absolute value of the temperature gradients (see Subsection \ref{cloudfactorsection}). But additional data points, including interactive data points in more elaborate models that incorporate physical processes influencing cloud cover at other latitudinal locations, might improve the climate sensitivity of the model considered here and add further insight into the dynamics of the polar jets.\\  

\textbf{Acknowledgements.} We thank the editor and anonymous reviewers for their careful readings, detailed corrections, and insightful suggestions.  We also thank Cristina Cadavid for discussions on albedo of solar radiation related to this work, Robert Fovell and  Jo\~ao Teixeira for helpful discussions and modeling suggestions, Aaron Donohoe for the data for Figure 3, and Matthew Levy and Paul Ryan for programming assistance.\\

\textbf{Funding and/or Conflicts of interests/Competing interests.} The authors have no relevant financial or non-financial interests to disclose. No funding was received for conducting this study.

\pagebreak

\begin{appendices}

\section{Cloud Factor Function Formula} \label{sec:A}

%
%
The formula for the cloud factor function is shown here when the extratropical maximum occurs at $\hat{\theta}$ latitude.  \begin{equation}\label{eq:cfc}
\scriptsize
C_{f}(\theta, \hat{\theta}) = 
\begin{cases}
0.000111111(\dfrac{30-\theta}{15}+1)\theta^2+0.001(\dfrac{\theta}{15}+1)(30-\theta)^2 & 0 \leq \theta \leq 30\\
\\

(0.1+0.2\Big(\dfrac{\theta-30}{\hat{\theta}-30}\Big))\Big(\dfrac{\hat{\theta}-\theta}{\hat{\theta}-30}\Big)^2 + (0.8+1.6\Big(\dfrac{\hat{\theta}-\theta}{\hat{\theta}-30}\Big))\Big(\dfrac{\theta-30}{\hat{\theta}-30}\Big)^2 & 30 \leq \theta \leq \hat{\theta}\\
\\

0.8 & \hat{\theta} \leq \theta \leq 90
\end{cases}
\end{equation}\\
The graph of Eq. \eqref{eq:cfc}  with $\hat{\theta} = 50^{\circ}$ is shown in Figure \ref{fig:cfs3}.

\section{Solution Methodology For The Initial Boundary Value Problem}\label{program}

\begin{figure}[H]
\center
\includegraphics[width=0.95\textwidth]{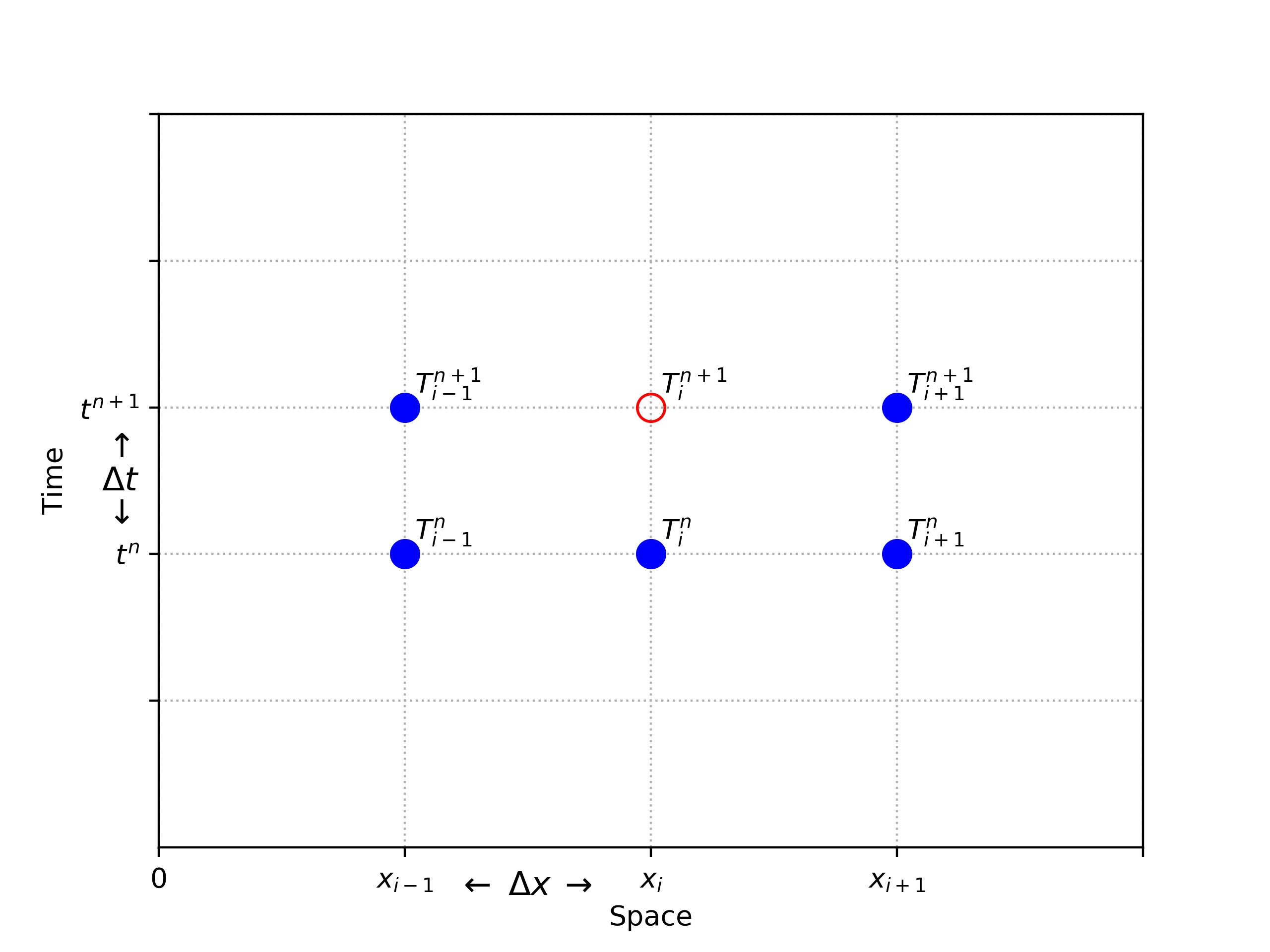}
\caption[Finite Difference Grid]{A schematic interpretation of the FDM approximation. The value of $T_{i}^{n+1}$ (hollow disk) requires the values of five neighbored points (solid disks).}
\label{fig:FDG}
\end{figure}

The initial boundary value problem (IBVP) \eqref{eq:ibvp} falls in the class of linear evolution problems for which various numerical methods have been developed. We have employed in this paper an \textit{implicit} finite difference method (FDM) based on the Crank-Nicholson scheme \cite{angermann,strikwerda}. This scheme has the desirable property of being inherently \textit{stable}. More specifically, we subdivide the spatial variable interval [0,1] uniformly in I subintervals $(x_i, x_{i+1})$, $i = 0, ..., I$ where $x_{i} = i\Delta x$; $\Delta x$ being the spatial step size that is set to be $10^{-3}$ (See Figure \ref{fig:FDG}). Similarly, we consider for the time variable t, the equidistant sequence  $t^{n} = n\Delta t$; $n = 0,1, ... , N$, where the time step $\Delta t$ is set to be 1 day and N is chosen large enough for the temperature to reach the asymptotic regime, i.e, the equilibrium of the solution of the IBVP\eqref{eq:ibvp}. For the simplicity of the publication, we introduce the auxiliary variable T to denote either the temperature of the atmospheric layer, $T_a$ or the temperature of the surface layer, $T_s$. We then approximate $T(x_i,t^n)$ by $T_i^n$ where  $T_i^n$ is the solution of the algebraic system resulting from the adopted finite difference scheme.
\\

The derivatives that occur in the IBVP \eqref{eq:ibvp} are approximated as follows. First, we have distributed the spatial derivative and then we have used the following \textit{second order} approximation,

\begin{equation}
\dfrac{\partial T}{\partial x}(x_{i},t^{n}) \approx \dfrac{T_{i+1}^{n} - T_{i-1}^{n}}{2\Delta x}
\label{eq:sdiv1}
\end{equation}

and

\begin{equation}
\dfrac{\partial^{2} T}{\partial x^{2}}(x_{i},t^{n}) \approx \dfrac{T_{i+1}^{n} - 2T_{i}^{n} + T_{i-1}^{n}}{\Delta x^{2}}.
\label{eq:sdiv2}
\end{equation}
\\
The first order time derivative is replaced by a \textit{second order} approximation using the Crank-Nicholson relations \cite{angermann,strikwerda}

\begin{equation}
\dfrac{\partial T}{\partial t}(x_{i},t^{n+\frac{1}{2}}) = \dfrac{1}{2}[\dfrac{\partial T}{\partial t}(x_{i},t^{n+1}) + \dfrac{\partial T}{\partial t}(x_{i},t^{n})]
\label{eq:CNS}
\end{equation}

and 

\begin{equation} 
\dfrac{\partial T}{\partial t}(x_i,t^{n+\frac{1}{2}}) \approx \dfrac{T_i^{n+1} - T_i^{n}}{\Delta t}
\label{eq:timediv}
\end{equation}

sequentially, IBVP\eqref{eq:ibvp} is then replaced by the following algebraic system, 

\begin{subequations}
\begin{equation}
\begin{aligned}
\beta(T_{a_i}^{n+1} - T_{a_i}^{n} ) &= \dfrac{1}{2}[\beta_{i}^{'}(T_{a_{i+1}}^{n+1} - 2T_{a_i}^{n+1} + T_{a_{i-1}}^{n+1}) - \beta_{i}^{''}(T_{a_{i+1}}^{n+1} -T_{a_{i-1}}^{n+1}) \\
& - (B_{up} + B_{out})T_{a_i}^{n+1} + B_{up}T_{s_i}^{n+1} + A_{up} - A_{out} \\
& + (1 - \alpha_{a_i}^n - \mathcal{T}_{sw_i}^n)(1+ \dfrac{\alpha_{g_i}^n\mathcal{T}_{sw_i}^n}{(1 - \alpha_{a_i}^n\alpha_{g_i}^n)})\dfrac{S_{0}s(x_{i})}{4} \\
& + \beta_{i}^{'}(T_{a_{i+1}}^{n} - 2T_{a_i}^{n} + T_{a_{i-1}}^{n}) - \beta_{i}^{''}(T_{a_{i+1}}^{n} - T_{a_{i-1}}^{n}) \\
& - (B_{up} + B_{out})T_{a_i}^{n} + B_{up}T_{s_i}^{n} + A_{up} - A_{out} \\
& + (1 - \alpha_{a_i}^n - \mathcal{T}_{sw_i}^n)(1+ \dfrac{\alpha_{g_i}^n\mathcal{T}_{sw_i}^n}{(1 - \alpha_{a_i}^n\alpha_{g_i}^n)})\dfrac{S_{0}s(x_{i})}{4}]
\end{aligned}
\label{eq:CNS1}
\end{equation}
\begin{equation}
\begin{aligned}
\gamma(T_{s_i}^{n+1} - T_{s_i}^{n}) &= \dfrac{1}{2}[\gamma_{i}^{'}(T_{s_{i+1}}^{n+1} - 2T_{s_i}^{n+1} + T_{s_{i-1}}^{n+1}) - \gamma_{i}^{''}(T_{s_{i+1}}^{n+1} -T_{s_{i-1}}^{n+1}) \\
& - B_{up}T_{s_i}^{n+1} + B_{up}T_{a_i}^{n+1} - A_{up} + \dfrac{(1 - \alpha_{g_i}^n)\mathcal{T}_{sw_i}^n}{(1 - \alpha_{a_i}^n\alpha_{g_i}^n)}\dfrac{S_{0}s(x_{i})}{4} \\
& + \gamma_{i}^{'}(T_{s_{i+1}}^{n} - 2T_{s_i}^{n} + T_{s_{i-1}}^{n}) - \gamma_{i}^{''}(T_{s_{i+1}}^{n} -T_{s_{i-1}}^{n}) \\
& - B_{up}T_{s_i}^{n} + B_{up}T_{a_i}^{n} - A_{up} + \dfrac{(1 - \alpha_{g_i}^n)\mathcal{T}_{sw_i}^n}{(1 - \alpha_{a_i}^n\alpha_{g_i}^n)}\dfrac{S_{0}s(x_{i})}{4}]
\end{aligned}
\label{eq:CNS2}
\end{equation}
\label{eq:CNS3}
\end{subequations}

\noindent where 
\begin{equation}
\begin{aligned}
\beta &= \dfrac{C_{a}}{\Delta t} & \gamma &= \dfrac{C_{s}}{\Delta t} \\
\beta_{i}^{'} &= \dfrac{C_{a}K_{a}(1 - x_i^{2})}{2a^2\Delta x^2} & \gamma_{i}^{'} &= \dfrac{C_{s}K_{s}(1 - x_i^2)}{2a^2\Delta x^2} \\
\beta_{i}^{''} &= \dfrac{2x_{i}C_{a}K_{a}}{4a^2\Delta x} & \gamma_{i}^{''} &= \dfrac{2x_{i}C_{s}K_{s}}{4a^2\Delta x}. 
\end{aligned}
\end{equation}
\\

A schematic interpretation or cone of dependance of the adopted FDM discretization is depicted in Figure \ref{fig:FDG}. It shows the implicit nature of this scheme. It also reveals that the evaluation of the temperature at the boundaries $T_{0}^{n}$(resp. $T_{I}^{n}$) requires the values of $T_{-1}^{n}$(resp. $T_{I+1}^{n}$). These ``fictitious" values are set to be $T_{-1}^{n} = T_{0}^{n}$ and $T_{I+1}^{n} = T_{I}^{n};\; n = 0, ... , N$. This choice results from the \textit{first order} approximation of the boundary condition, IBVP \eqref{eq:ibvp}.

Note that the algebraic system \eqref{eq:CNS3} can be expressed in a compact representation as follows,

\begin{equation}
\textbf{A}\textbf{T}^{n+1} = \textbf{B}\textbf{T}^{n} + \textbf{b}^n.
\label{eq:linsys}
\end{equation}

Where A and B are block diagonal matrices whose entries are explicitly given in equations C.1 - C.14, pages 88 - 92 in \cite{perillo}. The vector $\textbf{T}$ consists of the temperature values for the atmosphere followed by the surface temperature values. The components of the vector $\textbf{b}$ consist of all terms not linear in temperature.  The linear system \eqref{eq:linsys} is solved using LAPACK package (\textit{routine -gesv})\cite{gesv} that is based on LU type decomposition \cite{golub}.\\

The temperature gradients reported in Figures \ref{sumfigures}, \ref{sumfigures2}, and \ref{sumfigures3} have been evaluated with the software package (\textit{numpy.gradient})\cite{grad}. This routine computes the gradient using second order accurate central differences in the interior points and either first or second order accurate one-side differences at the boundaries.

\newpage
\section{Quasi-Periodic Oscillations of Polar Jet}\label{quasiperiodic}
\begin{table}[htp]
\caption{Latitudes of Polar Jet defined by $\max|\partial/\partial \theta (T_a + T_s)|$\newline
for $t^n \in [8725, 8760]$ (in days) with $A_{\text out}$ in Wm$^{-2}$. The last two rows give means and standard deviations for each column.}  
\vspace{.5cm}
\resizebox{\textwidth}{!}{
\begin{tabular}{||c |c|c|c|c|c|c|c|c|c||}
\hline
$\mathbf{A_{\text out}=211}$ & $\mathbf{A_{\text out}=210}$ & $\mathbf{A_{\text out}=209}$ &$\mathbf{A_{\text out}=208}$&$\mathbf{A_{\text out}=207}$&$\mathbf{A_{\text out}=206}$&$\mathbf{A_{\text out}=205}$&$\mathbf{A_{\text out}=204}$&$\mathbf{A_{\text out}=203}$&$A\mathbf{_{\text out}=202}$\\
\hline
64.027 & 64.027 & 49.025 & 64.027 & 45.480& 43.079& 42.766& 41.913& 41.224& 39.941\\
64.027 & 50.624 & 48.851 & 47.223 & 45.072& 43.709& 42.144& 41.300& 40.617& 40.542\\
64.027 & 64.027 & 64.027 & 47.138 & 45.398& 43.079& 42.766& 41.913& 41.224& 39.941\\
52.279 & 64.027 & 48.938 & 47.054 & 44.991& 43.709& 42.144& 41.300& 40.617& 40.542\\
64.027 & 50.714 & 64.027 & 64.027 & 45.398& 43.079& 42.766& 41.913& 41.224& 39.941\\
64.027 & 64.027 & 49.025 & 47.307 & 44.910& 43.709& 42.144& 41.300& 40.617& 40.542\\
64.027 & 50.624 & 48.851 & 47.138 & 64.027& 43.079& 42.766& 41.913& 41.224& 39.941\\
64.027 & 64.027 & 64.027 & 47.138 & 45.643& 43.709& 42.144& 41.300& 40.617& 40.542\\
52.279 & 50.624 & 48.938 & 46.970 & 45.316& 43.079& 42.766& 41.913& 41.224& 39.941\\
64.027 & 64.027 & 48.851 & 64.027 & 45.561& 43.709& 42.144& 41.300& 40.617& 40.542\\
64.027 & 64.027 & 64.027 & 47.223 & 45.154& 43.079& 42.766& 41.913& 41.224& 39.941\\
64.027 & 50.714 & 48.938 & 47.138 & 45.480& 43.709& 42.144& 41.300& 40.617& 40.542\\
64.027 & 64.027 & 64.027 & 47.054 & 45.072& 43.079& 42.766& 41.913& 41.224& 39.941\\
64.027 & 50.624 & 49.025 & 64.027 & 45.398& 43.709& 42.144& 41.300& 40.617& 40.542\\
52.279 & 64.027 & 48.851 & 47.223 & 44.991& 43.079& 42.766& 41.913& 41.224& 39.941\\
64.027 & 64.027 & 64.027 & 47.138 & 64.027& 43.709& 42.144& 41.300& 40.617& 40.542\\
64.027 & 50.714 & 48.938 & 47.054 & 45.643& 43.079& 42.766& 41.913& 41.224& 39.941\\
64.027 & 64.027 & 64.027 & 64.027 & 45.316& 43.709& 42.144& 41.300& 40.617& 40.542\\
64.027 & 50.714 & 49.025 & 47.307 & 45.561& 43.079& 42.766& 41.913& 41.224& 39.941\\
52.279 & 64.027 & 48.851 & 47.138 & 45.154& 43.709& 42.144& 41.300& 40.617& 40.542\\
64.027 & 50.624 & 64.027 & 47.138 & 45.480& 43.079& 42.766& 41.913& 41.224& 39.941\\
64.027 & 64.027 & 48.938 & 46.970 & 45.072& 43.709& 42.144& 41.300& 40.617& 40.542\\
64.027 & 64.027 & 64.027 & 64.027 & 45.398& 43.079& 42.766& 41.913& 41.224& 39.941\\
64.027 & 50.714 & 49.025 & 47.223 & 44.991& 43.709& 42.144& 41.300& 40.617& 40.542\\
64.027 & 64.027 & 48.938 & 47.138 & 45.316& 43.079& 42.766& 41.913& 41.224& 39.941\\
52.279 & 50.624 & 64.027 & 46.970 & 44.910& 43.709& 42.144& 41.300& 40.617& 40.542\\
64.027 & 64.027 & 48.938 & 64.027 & 64.027& 43.079& 42.766& 41.913& 41.224& 39.941\\
64.027 & 50.624 & 48.851 & 47.307 & 45.643& 43.709& 42.144& 41.300& 40.617& 40.542\\
64.027 & 64.027 & 64.027 & 47.138 & 45.316& 43.079& 42.766& 41.913& 41.224& 39.941\\
64.027 & 64.027 & 48.938 & 47.054 & 45.480& 43.709& 42.144& 41.300& 40.617& 40.542\\
52.279 & 50.714 & 64.027 & 64.027 & 45.154& 43.079& 42.766& 41.913& 41.224& 39.941\\
64.027 & 64.027 & 49.025 & 47.307 & 45.398& 43.709& 42.144& 41.300& 40.617& 40.542\\
64.027 & 50.624 & 48.851 & 47.138 & 45.072& 43.079& 42.766& 41.913& 41.224& 39.941\\
64.027 & 64.027 & 64.027 & 47.138 & 45.398& 43.709& 42.144& 41.300& 40.617& 40.542\\
64.027 & 50.624 & 48.938 & 46.970 & 44.910& 43.079& 42.766& 41.913& 41.224& 39.941\\
64.027 & 64.027 & 64.027 & 64.027 & 64.027& 43.709& 42.144& 41.300& 40.617& 40.542\\
\hline\hline
Mean: 61.937 & Mean: 58.337 & Mean: 54.762 & Mean: 51.152 & Mean: 47.340& Mean: 43.397& Mean: 42.452& Mean: 41.603& Mean: 40.917& Mean: 40.244\\
\hline\hline
STD: 4.510 & STD: 6.641 &STD: 7.385 &STD: 7.224 &STD: 5.867&STD: 0.317&STD: 0.312&STD: 0.308&STD: 0.305&STD: 0.302\\
\hline\hline
\end{tabular}
}
\end{table}%

\end{appendices}

\end{document}